\documentclass[twocolumn,longbibliography,10pt,aps,pra]{revtex4-2}

\usepackage{amsmath,dsfont}
\usepackage{graphicx,xcolor}
\usepackage{mathtools}
\usepackage{hyperref}
\usepackage{lmodern}

\hypersetup{
    colorlinks,
    linkcolor=blue,
    citecolor=blue,
    urlcolor=blue
}

\DeclareMathOperator{\sech}{sech}
\DeclareMathOperator{\trian}{tri}
\DeclareMathOperator{\rect}{rect}
\DeclareMathOperator{\sign}{sgn}

\begin{document}

\title{Shaping the~\boldmath{$g^{(2)}$} autocorrelation and photon statistics}

\author{Ivo Straka}
\email{straka@optics.upol.cz}
\author{Miroslav Je\v{z}ek}
\affiliation{Department of Optics, Faculty of Science, Palack\'{y} University, 17. listopadu 12, 771~46, Olomouc, Czechia}

\begin{abstract}
We propose a~method of arbitrarily shaping and scaling the~temporal intensity correlations of an~optical signal locally, avoiding periodic correlations. We demonstrate our approach experimentally using stochastic intensity modulation. We also analyze and simulate shaping both temporal correlations and photon statistics that are fully specified by the~user. We show that within the~confines of monotony and convexity, the~temporal correlations are independent of photon statistics and can take any shape.
\end{abstract}
\maketitle

\section{Introduction}

This article focuses on the temporal correlations of optical intensity $I(t)$, conventionally described by the~second-order coherence function
\begin{equation}\label{eq:g2def}
g^{(2)}(\tau) \coloneqq \frac{\langle I(t) I(t+\tau) \rangle}{\langle I(t) \rangle^2}
\end{equation}
which we refer to as intensity autocorrelation. In the~quantum realm, it is defined by the~means of normally ordered photon-number operators \cite{Loudon}
\begin{equation}
g^{(2)}(\tau) \coloneqq \frac{\langle \vcentcolon \mathrel{\hat{n}(t) \hat{n}(t+\tau)} \vcentcolon \rangle}{\langle \hat{n}(t) \rangle^2}.
\end{equation}

This article proposes methods of shaping the~autocorrelation $g^{(2)}(\tau)$ in an~arbitrary way. A~straightforward approach would be modulating a~pulse train coming from a~laser, however pulsed signals have periodic autocorrelation. Our goal is to emulate fluctuating signals exhibiting $g^{(2)}(0) > 1$ that approaches unity in the~limit of $\tau \to \infty$. Any significant deviation from unity is therefore local.

The~modulation of optical intensity in time can be used to generate pseudothermal light. A~common technique is collecting speckles from light diffused by a~rotating ground glass plate \cite{Martienssen2005Jul}. Chaotic light can be also generated by amplified spontaneous emission \cite{Boitier2009Apr} or direct laser modulation \cite{Nazarathy1989Jan}. The~resulting photon bunching of $g^{(2)}(0) \geq 2$ can be exploited in classical ghost imaging~\cite{Bennink2002Aug,Gatti2004Aug,Zhou2017May}, two-photon-excited fluorescence \cite{Jechow2013Dec}, optical time-domain reflectometry \cite{Wang2008Jul}, and laser ranging \cite{Lin2004Dec}.

Shaping the~autocorrelation---and thus the~spec\-trum---of a~signal is a~problem that has been addressed in other fields of signal processing and simulation using various methods \cite{Kim2010,Mueller2012}. A~prominent case is transforming an~electronic Gaussian noise by a~linear filter and a~nonlinear function to obtain a~given spectrum and marginal distribution \cite{Liu1982}. An~alternate solution was proposed in terms of reshuffling of signal samples \cite{Hunter1983Jun}.

In the~first part of this article, we present a~method of shaping the~autocorrelation that exploits stochastic pulse superposition. The~spectral properties of such random pulses have been studied in communication theory \cite{Leneman1967Sep}; we make use of these properties from the~perspective of signal generation.

In the~second part, we utilize another method of stochastic modulation---random switching of intensity levels that was proposed in Ref.~\cite{Pandey2014Jun}---to shape both temporal correlations and photon statistics. We combine this approach with the~photon-statistics-generation algorithm proposed in Ref.~\cite{Straka2018Apr}. Using numerical simulations, we demonstrate the~generation of arbitrarily specified $g^{(2)}(\tau)$ and photon-number distribution. We also analyze the~properties and limitations of the~method.

\section{Arbitrary autocorrelation}
\label{sec:arbitrary_autocorrelation}

\subsection{Method}

The~basic premise is that the~intensity signal $I(t)$ consists of a~random Poissonian sequence of non-negative integrable hill functions $h(t)$. Mathematically, the~signal is expressed as a~sum
\begin{equation}
I(t) = \sum_i h(t - t_i).
\end{equation}
The~sequence of positions in time $\{t_i\}$ follows a~homogeneous Poisson point process with a~mean rate $\lambda$. This means that every delay $\Delta t_i = t_{i+1}-t_i$ is a~realization of a~random variable with a~probability density
\begin{equation}
p(\Delta t) = \lambda\exp(-\lambda \Delta t).
\end{equation}
The~hill function is chosen so that it has a~specific cross-correlation $C_{\mathrm{norm}}(\tau)$, while its $L^1$ norm $\|h\|$ may be arbitrary. The~terms are defined as
\begin{align}\label{eq.hnormdef}
\|h\| &\coloneqq \int_{-\infty}^\infty h(t) \mathrm{d}t,	\\\label{eq.Cdef}
C_{\mathrm{norm}}(\tau) &\coloneqq \frac{1}{\|h\|^2}\int_{-\infty}^\infty h(t+\tau) h(t) \mathrm{d}t,
\end{align}
where the~cross-correlation observes $\int_{-\infty}^\infty C_{\mathrm{norm}}(\tau)\mathrm{d}\tau = 1$. Then, the~intensity autocorrelation of $I(t)$ becomes

\begin{equation}\label{eq:g2}
g^{(2)}(\tau) = 1 + \frac{1}{\lambda}C_{\mathrm{norm}}(\tau).
\end{equation}
The~proof is given in Appendix \ref{ap:derivation_hills}. The~autocorrelation shape is therefore given solely by the~shape of $h(t)$, the~amplitude of which may be arbitrary. The~$g^{(2)}(0)$ and the~scale of the~whole autocorrelation can be set by the~rate $\lambda$.

Given a~specific $g^{(2)}(\tau)$, one can obtain the~hill function using Fourier transforms $\mathcal{F}$ (see Appendix \ref{ap:hill_inverse}),
\begin{align}
\widehat{C}(\nu) &= \mathcal{F}\left[g^{(2)}(\tau)-1\right], \\
\label{eq:hill_formula}
h(t) &= \mathcal{F}^{-1}\left[ \sign(\nu) \sqrt{\widehat{C}(\nu)} \right].
\end{align}
For any $g^{(2)}(\tau)$ consistent with \eqref{eq:g2def}, the~function $\widehat{C}(\nu)$ is real and non-negative by virtue of the~Wiener-Khinchin theorem. The~sign function in \eqref{eq:hill_formula} is any piecewise function yielding $\pm 1$. The~constraint of this method is that the~ $g^{(2)}(\tau)$ needs to be given so that there exists a~sign function that results in a~non-negative $h(t)$.

\paragraph{Superposition}
A generalization of this approach allows generating a~superposition of multiple shapes by randomly alternating between hill functions. Suppose a~set of hill functions $\{h_n(t)\}$ with respective cross-correlations $C^{(n)}_{\mathrm{norm}}(\tau)$ appearing with probabilities $p_n$, but sharing the~same mean frequency $\lambda$. The~calculation of $g^{(2)}(\tau)$ is a~straightforward extension, yielding
\begin{equation} \label{eq:superposition}
g^{(2)}_\text{mul} (\tau) = 1+\frac{1}{\lambda} \frac{\sum_n p_n C^{(n)}_{\mathrm{norm}}(\tau)}{\left(\sum_n p_n \|h\|_n\right)^2}.
\end{equation}

\paragraph{Background}
If we consider a~background offset $I_\mathrm{bg}$ so that
$I(t) = \sum_i h(t - t_i) + I_\mathrm{bg}$,
the~mean value of the~signal is $\langle I(t) \rangle = \lambda \|h\| + I_\mathrm{bg}$ and the~signal-to-noise ratio between $\|h\|$ and $I_\mathrm{bg}$ becomes a~factor that scales down the~intensity autocorrelation:
\begin{equation}\label{eq.g2bg}
g^{(2)}_\text{bg}(\tau) = 1 + \frac{C_{\mathrm{norm}}(\tau)}{\lambda \left( 1+ \frac{I_\mathrm{bg}}{\lambda \|h\|} \right)^2}.
\end{equation}
The~maximum bunching $g^{(2)}(0)$ possible is achieved for $\lambda = I_\mathrm{bg}/\|h\|$.

\paragraph{Non-overlapping hills}
In a~particular experimental realization---such as presented here---it may not be possible to generate a~mathematical superposition of the~hill functions. For example, a~function generator can only produce one pulse at a~time and so the~hills cannot overlap. In such cases, the~hill function is defined on a~finite support $t \in (-t_0/2, t_0/2)$ and the~minimum distance between the~hills' peaks is $t_0$. Consequently, the~cross-correlation $C_{\mathrm{norm}}(\tau)$ is only nonzero for $\tau \in (-t_0, t_0)$, and the~unity term in \eqref{eq:g2} becomes locally distorted. The~expression changes to (see Appendix \ref{ap:non-overlapping})

\begin{align}\label{eq:non_overlap}
g^{(2)}_\text{n-ol}(\tau) &= \left ( \frac{1}{\lambda} + t_0 \right) C_{\mathrm{norm}}(\tau) + \left( 1 + \lambda t_0 \right) \\\nonumber
&\qquad \times \sum_{k=0}^{\left\lfloor |\tau|/t_0 \right\rfloor} \int_{t=0}^{|\tau| - k t_0} \Big[ C_{\mathrm{norm}} \left(\tau + t + \left( k+1 \right) t_0 \right) \\\nonumber
&\qquad\qquad + C_{\mathrm{norm}} \left(\tau - t - \left(k+1 \right) t_0 \right) \Big] \lambda^k \frac{t^k}{k!} \mathrm{e}^{-\lambda t} \mathrm{d} t.
\end{align}

\subsection{Results}

\begin{figure}
	\centering
	\includegraphics[width=\linewidth]{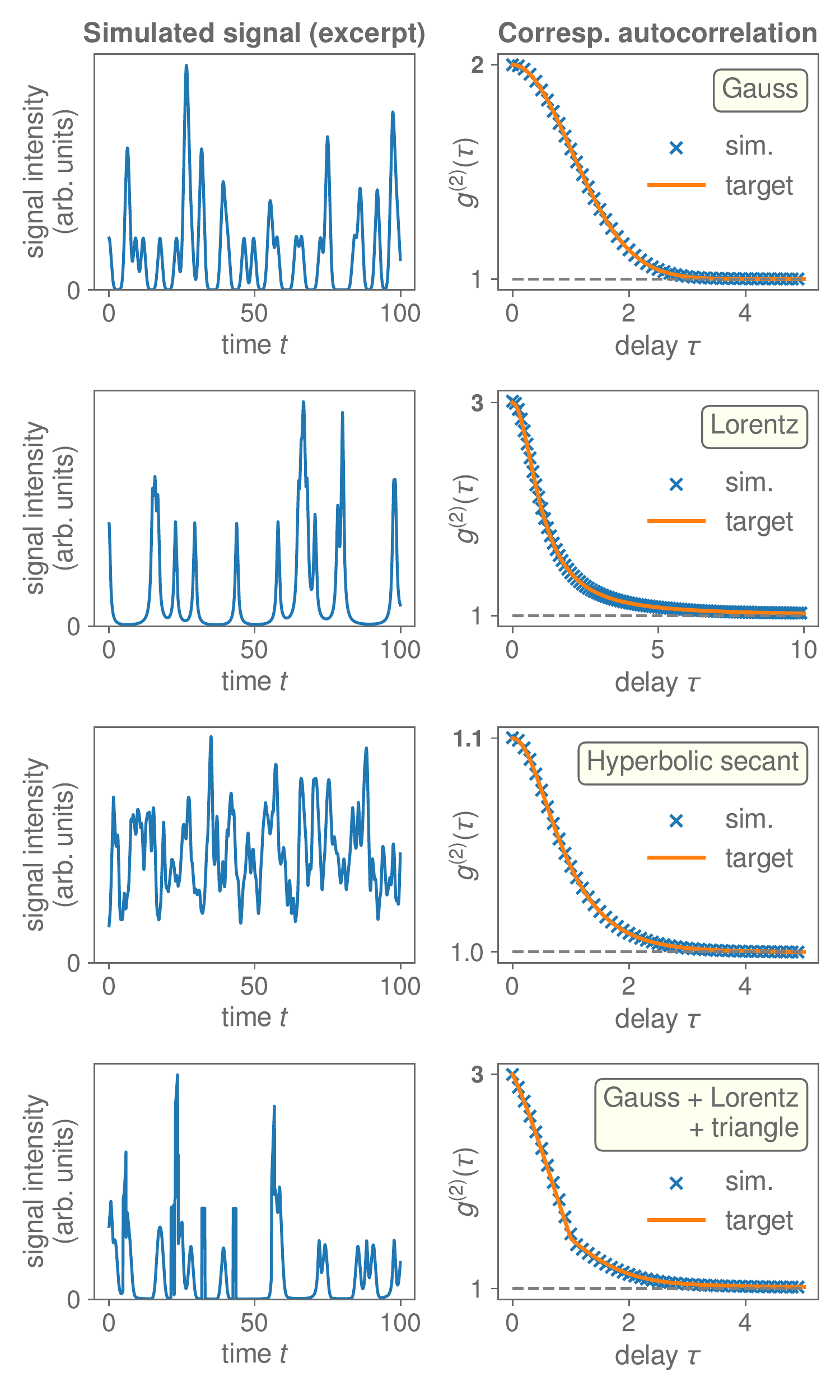}
	\caption{Monte-Carlo simulations of signals with various prescribed autocorrelation shapes and values of $g^{(2)}(0)$; the source code is available online \cite{Code}. The~shapes and parameters are listed in Table \ref{tab1}.}
	\label{fig:sim1all}
\end{figure}

\begin{table*}
	\centering
	\begin{ruledtabular}
		\begin{tabular}{lcccc}
			Shape & $C_{\mathrm{norm}}(\tau)$ & $h(t)$ & $g^{(2)}(0)$ & $\lambda$	\\\hline
			Gauss & $\frac{1}{\sqrt{2\pi}}\mathrm{e}^{-\tau^2/2}$ & $\frac{1}{\sqrt{\pi}}\mathrm{e}^{-t^2}$  & 2 & $\frac{1}{\sqrt{2\pi}}$	\\
			Cauchy--Lorentz & $\frac{1}{\pi (1+\tau^2)}$ & $\frac{2}{\pi(1+4t^2)}$ & 3 & $\frac{1}{2\pi}$	\\
			Hyperb.\@ secant & $\frac{1}{2}\sech{\frac{\pi \tau}{2}}$	& $\mathcal{F}^{-1}\left[ \sqrt{\sech(2\pi\nu)}\right] $ & $1.1$ & 5	\\
			\begin{tabular}{l}
				Gauss \\
				+ Lorentz \\
				+ triangle
			\end{tabular}
			&
			\begin{tabular}{l}
				$\frac{1}{3} \frac{1}{\sqrt{2\pi}} \mathrm{e}^{-\tau^2/2}$ \\
				$+\frac{1}{3} \frac{1}{\pi (1+\tau^2)}$	\\
				$+\frac{1}{3} \trian(-1,1)$
			\end{tabular}
			& 
			\begin{tabular}{l}
				$h_1(t) = \frac{1}{\sqrt{\pi}} \mathrm{e}^{-t^2}$ \\
				$h_2(t) = \frac{2}{\pi(1+4t^2)}$	\\
				$h_3(t) = \rect(-\frac{1}{2},\frac{1}{2})$
			\end{tabular}
			& 3 &
			$\frac{1}{6}\left( 1 + \frac{1}{\pi} + \frac{1}{\sqrt{2\pi}} \right) $
			\\
			Nonmonotone &
			\begin{tabular}{l}
				$\frac{3-|\tau|}{18}\left[ 2+\cos(2\pi \tau) \right]$ \\ $+\frac{1}{12\pi}\sin(2\pi|\tau|)$
				for $|\tau| < 3$
			\end{tabular}
			&
			\begin{tabular}{l}
				$\frac{1}{2}-\frac{1}{2}\cos(2\pi t)$ \\
				for $0<t<3$
			\end{tabular}
			&  & \\ 
		\end{tabular}
	\end{ruledtabular}
	\caption{Table of autocorrelation shapes used in Fig.~1. Figure~2 uses the~same shapes and the~nonmonotone shape, which is defined at the~bottom. The~functions $\rect()$ and $\trian()$ are rectangular and triangular functions of unit height and given edges.}
	\label{tab1}
\end{table*}

\begin{figure}
	\includegraphics[width=\linewidth]{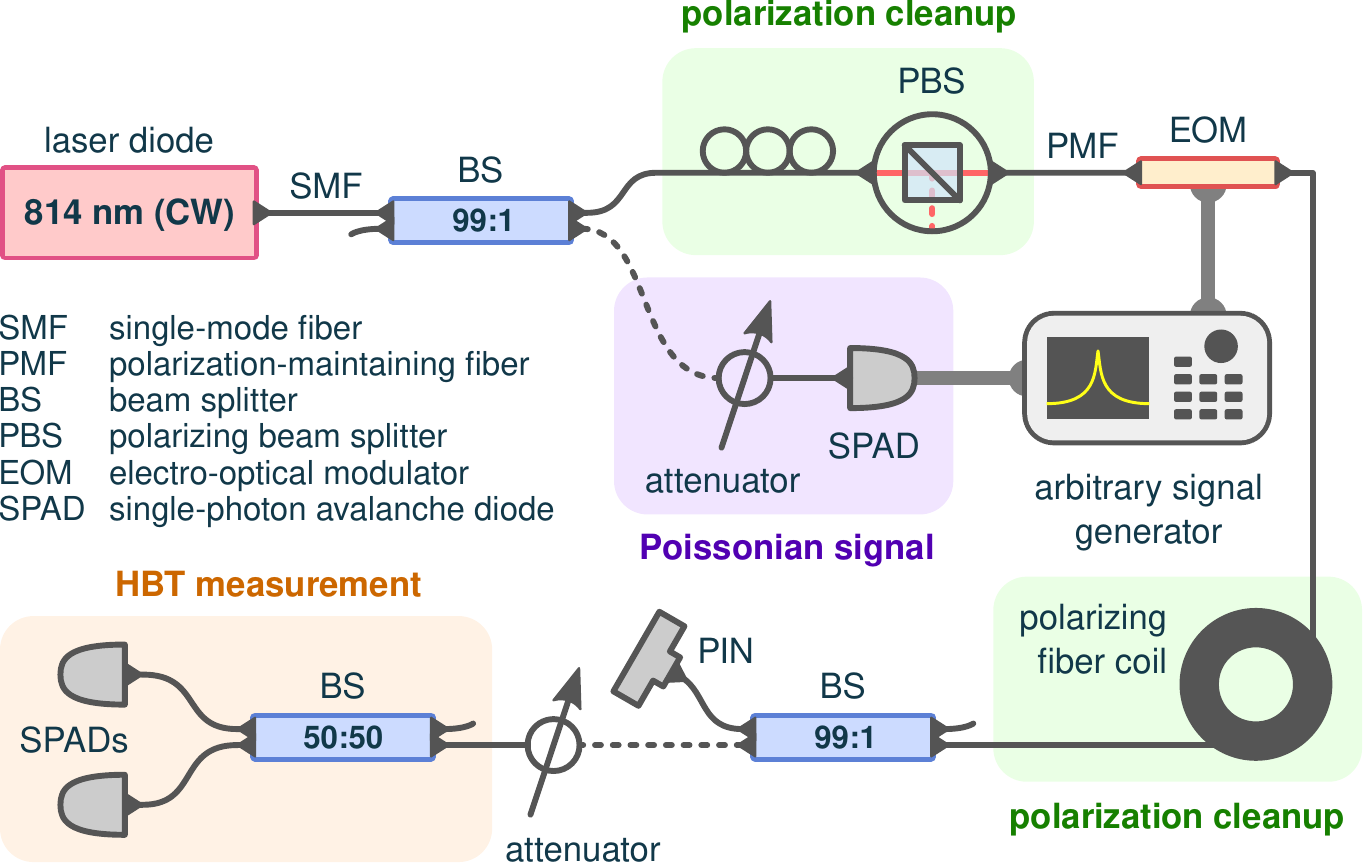}
	\caption{The~measurement scheme with the~data given in Fig.~\ref{fig:data}.}
	\label{fig:setup}
\end{figure}

\begin{figure}
	\centering
	\includegraphics[width=.75\linewidth]{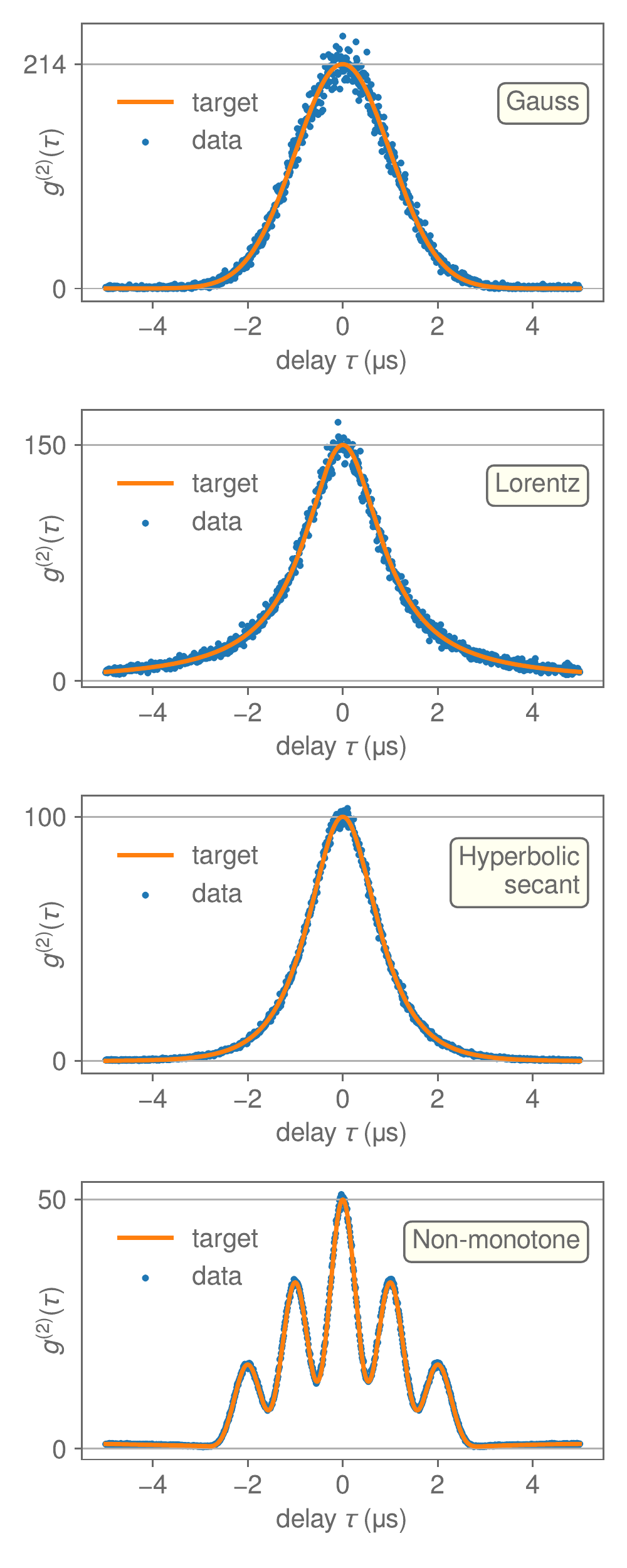}
	\caption{The~autocorrelation functions measured with the~HBT configuration. The~good match between the~prescribed orange curves and the~data was achieved by generating the~voltage pulses of the~appropriate shape, by tuning the~Poissonian trigger rate $\lambda$, and by setting the~EOM bias voltage as precisely as possible. All data and code are available online \cite{Code}.}
	\label{fig:data}
\end{figure}

The~method of stochastic signal generation is simulated in Fig.~\ref{fig:sim1all}. For illustrative purposes, the~signals are shown for both high and low intensities $\lambda$, respectively corresponding to low and high values of $g^{(2)}(0)$. A~mixture of three shapes corresponding to \eqref{eq:superposition} is given in the~last row.

Experimental evidence was obtained using the~measurement setup depicted in Fig.~\ref{fig:setup}.
The~source was a~single-mode-fiber-coupled (SMF) continuous-wave laser diode at 814 nm. The~optical signal was coupled into a~polarization-maintaining fiber (PMF) using a~manual polarization controller and a~polarizing beam splitter (PBS). The~intensity of the~signal was modulated by an~integrated electro-optical Mach-Zehnder modulator (EOM; EOSpace). The~modulator was driven by an~arbitrary signal generator (Tektronix) programed to emit a~voltage pulse upon each triggering event, which was fed into the~fast radio-frequency input of the~EOM, having the~shape
\begin{equation}
V(t) = \frac{V_\pi}{\pi}\arccos \left[ 1-2h(t)\right],
\end{equation}
where the~hill is scaled so that $h(t) \in [0,1]$, and $V_\pi$ is the~EOM half-wave voltage. The~induced transmission modulation is equal to $h(t)$ due to the~cosine response of the~EOM. We assume that the~constant bias voltage of the~EOM is set precisely to the~zero-level transmission.
The~triggering came from a~single-photon avalanche diode (SPAD) illuminated by a~constant intensity that was set to a~mean count rate $\lambda$. This served as the~source of Poissonian events. The~output was split between a~\textit{p-i-n} diode and a~single-photon Hanbury-Brown--Twiss (HBT) measurement \cite{Loudon}. The~\textit{p-i-n} served for live monitoring of the~output and the~HBT configuration measured the~$g^{(2)}(\tau)$. For all measurements, the~minimum pulse-to-pulse delay was $t_0 \geq 3$~\textmu{}s. It means that the~triggering SPAD dead time of 29~ns could be safely neglected. The~data from the~SPADs were acquired by a~81-ps time-tagging unit (qutools), the~coincidence window for the~HBT autocorrelation measurement was 10~ns, and the~measurement time was 100~s for each set.

The~results of the~HBT measurement are shown in Fig.~\ref{fig:data}. The~Gaussian autocorrelation was set to the~highest possible amplitude achievable for the~Gaussian, $g^{(2)}(0) = 214$, which corresponds to a~modulation dynamic range 30.8~dB; the~others were set to round values 50, 100, and 150. The~modulation scheme corresponds to the~non-overlapping scenario with the~expected autocorrelation given by \eqref{eq:non_overlap}, where the~effect of non-unity background is small compared to the~overall amplitude \cite{Code}.

The~limiting factor for autocorrelation speed is the~225-MHz bandwidth of the~arbitrary signal generator. The~limiting factor for the~amplitude is the~dynamic range of the~modulation. It is reduced by thermally induced phase drifts in the~EOM, by the~polarization extinction ratio, by the~EOM interferometric visibility, and by the~detectors' dark count rate. Consequently, the~dynamic range determines the~maximum bunching $g^{(2)}(0)$ achievable for each autocorrelation shape (Gaussian, Lorentz, etc.), as given by the~optimal rate $\lambda = I_\mathrm{bg}/\|h\|$ in \eqref{eq.g2bg}. For a more detailed discussion of the shapes and maximum bunching, see Appendix~\ref{ap:derivation_hills}.

\section{Photon statistics and autocorrelation}
\label{sec:g2_and_stat}

There is a~possibility of tailoring both temporal autocorrelation and photon (intensity) statistics simultaneously and independently. The~technique relies on random switching between intensity levels. This approach was proposed by Pandey \emph{et al}.~\cite{Pandey2014Jun} who demonstrated the~generation of gamma-distributed intensity and exponential autocorrelation shapes. In order to reach arbitrary classical photon statistics, we employ an~inversion algorithm proposed in our earlier work \cite{Straka2018Apr}. We present an~analysis of the~limitations and possibilities of the~resulting approach and we perform numerical simulations \cite{Code}.

The~optical signal consists of a~sequence of intensity levels $\{I_i\}$, each being held for a~respective time period $\{\Delta t_i\}$. Each $I_i$ and $\Delta t_i$ is an~independent realization of the~corresponding random variable, following the~probability density functions $p(I)$ and $p(\Delta t)$.

The~autocorrelation is then given as \cite{Pandey2014Jun} (see Appendix~\ref{ap:derivation_levels})
\begin{align}
g^{(2)}(\tau) &= 1 + \frac{g^{(2)}(0) - 1}{\langle \Delta t \rangle}\int_\tau^\infty p(\Delta t)(\Delta t - \tau) \mathrm{d}\Delta t,\\\label{eq:g20}
g^{(2)}(0) &= \frac{\langle I^2 \rangle}{\langle I \rangle^2}.
\end{align}
The~averaging terms $\langle \cdot \rangle$ are always taken over the~corresponding distributions---$p(I)$ or $p(\Delta t)$. The~bunching factor (\ref{eq:g20}) is solely given by intensity statistics---as it needs to be---while the~temporal shape is given by $p(\Delta t)$. To establish the~set of possible $g^{(2)}$ shapes, let us look at the~derivatives
\begin{align}\label{eq:g2_derivative1}
\frac{\mathrm{d}g^{(2)}(\tau)}{\mathrm{d}\tau} &= -\frac{g^{(2)}(0)-1}{\langle \Delta t \rangle} \int_\tau^\infty p(\Delta t) \mathrm{d}\Delta t \quad\leq 0,	\\
\label{eq:g2_derivative2}
\frac{\mathrm{d}^2 g^{(2)}(\tau)}{\mathrm{d}\tau^2} &= \frac{g^{(2)}(0)-1}{\langle \Delta t \rangle} p(\tau) \quad\geq 0.
\end{align}
The~above inequalities follow from $g^{(2)}(0) > 1$ and the~non-negativity of $p(\Delta t)$. Both inequalities mean that the~class of practicable autocorrelation functions is restricted to monotonically non-increasing convex functions for which $g^{(2)}(0) > 1$ and the~limit $\lim_{\tau \to \infty} g^{(2)}(\tau) = 1$. As is proven in Appendix \ref{ap:existence_levels}, \emph{any} such autocorrelation can be generated by this method.

These conditions are more restrictive than the~hill-superposition approach above, as for example all functions in Figs.~\ref{fig:sim1all} and \ref{fig:data} are nonconvex and therefore inaccessible. However, we gain the~possibility to shape the~photon statistics independently.

Photon statistics is given by the~integrated intensity
\begin{equation}
W(t) \coloneqq \int_t^{t+T} I(t') \mathrm{d}t',
\end{equation}
taken over a~time interval $T$ \cite{MandelWolf}. Let us consider the~units of $I(t)$ to be the~average number of photons per time, meaning that $W$ is dimensionless. The~number of photons $n$ detected inside a~time interval follows the~probability distribution given by Mandel's formula \cite{MandelWolf}
\begin{equation}\label{eq:Mandel}
p(n) = \int_0^\infty p(W) \frac{W^n}{n!}\mathrm{e}^{-W} \mathrm{d}W.
\end{equation}
In order to have control over the~photon statistics, the~distribution $p(W)$ needs to follow $p(I)$, ideally when $W(t)=T \times I(t)$, implying $p(W) = p(I)/T$. This can be achieved either in the~limit of a~short time window $T \to 0$, or if the~intensity switching is synchronized with the~time windows. The~latter implies that the~window length $T$ becomes an~elementary time unit of the~modulation, and the~autocorrelation is restricted to a~piecewise linear function with the~resolution $T$. This case corresponds to our discrete numerical simulations.

Following the~approach presented in Ref.~\cite{Straka2018Apr}, we take a~finite set $\{W_k\}$ and construct a~matrix $A_{nk} = \exp(-W_k) W_k^n/n!$. Each $W_k$ has a~corresponding probability $P_k$. The~Mandel formula (\ref{eq:Mandel}) then becomes
\begin{equation}\label{eq:Mandel_discrete}
p(n) = \sum_k A_{nk}P_k,
\end{equation}
where the~vector $P_k$ needs to be solved for a~given photon statistics $p(n)$. Suitable sampling and non-negative least squares were shown to work for classical photon statistics \cite{Straka2018Apr,Code}. The~solution for $p(I)$ is discrete: $\Pr[I = W_k/T] = P_k$.

Let us now explicitly outline the~problem inversion and generation of the~random samples of $I$ and $\Delta t$. We are given the~target $g^{(2)}(\tau)$ and $p(n)$. If $p(n)$ is a~classical photon statistics, then by definition it is achievable by sampling a~certain $p(W)$. If the~technical limitations of the~experiment only allow discrete intensities $\{W_k\}$, we perform a~non-negative least-squares inversion of \eqref{eq:Mandel_discrete} to obtain $P_k$. In each modulation step, we can randomly choose the~intensity $I_k=W_k/T$ using a~uniformly distributed random floating-point number $r \in [0,1)$ by finding the~smallest $k$ such that $\sum_{i \leq k} P_i > r$ \cite{Code}. This way, $I_k$ follows the~distribution $P_k$.

To shape the~autocorrelation, we establish a~survival function using the~first derivative $g'(\tau) \coloneqq \mathrm{d}g^{(2)}(\tau)/\mathrm{d}\tau$,
\begin{align}
S(\tau) = \frac{g'(\tau)}{g'(0)}.
\end{align}
In each step, we obtain the~hold-off time $\Delta t$ from a~uniformly distributed number $r' \in (0,1]$ using the~inverse function that always exists,
\begin{equation}
\Delta t \leftarrow S^{-1}(r').
\end{equation}
Consequently, $\Delta t$ follows the~probability density $p(\Delta t)$. The~proof of existence and derivations are given in Appendix \ref{ap:existence_levels}.

\begin{figure}
\centering
\includegraphics[width=\linewidth]{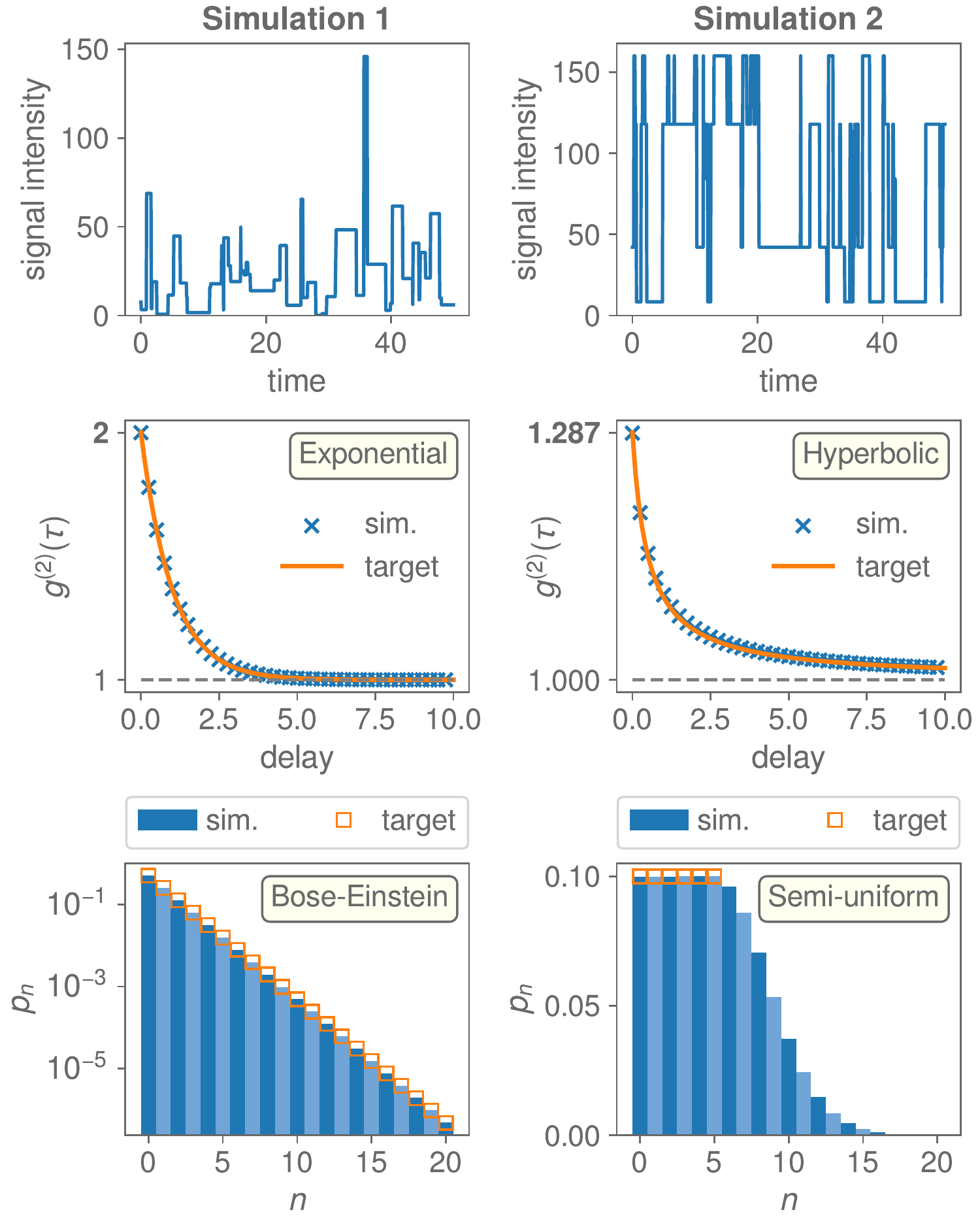}
\caption{Numerical simulations of signals having prescribed autocorrelations and photon statistics. Signal excerpts are shown in the~top row; autocorrelations in the~middle row; photon statistics per window $T=0.05$ are shown at the~bottom. The code is available online \cite{Code}.}
\label{fig:g2_stat_simulations}
\end{figure}

\begin{table}
\centering
\begin{ruledtabular}
\begin{tabular}{rcc}
& $g^{(2)}(\tau)$ & $p_n$ \\\hline
Simulation 1 & $1 + \mathrm{e}^{-|\tau|}$ & $0.5^{n+1}$	\\
Simulation 2 & $1+0.287\frac{1}{2|\tau|+1}$ & $p_{0,1,\dots,5} = 0.1$ \\
\end{tabular}
\end{ruledtabular}
\caption{Table of autocorrelation shapes and photon-number distributions used in Fig.~4.}
\label{tab2}
\end{table}

Figure~\ref{fig:g2_stat_simulations} shows two numerically simulated signals that follow certain prescribed autocorrelations and photon statistics. Both signals observe their corresponding $p(\Delta t)$, but have different approaches to $p(I)$. Signal~1 takes samples from a~continuous exponential distribution, whereas signal~2 follows a~discrete solution of \eqref{eq:Mandel_discrete} for a~partially uniform photon statistics $p(n) = 0.1$ $\forall n \leq 5$. The~corresponding prescriptions are given in Table~\ref{tab2}.

\section{Conclusion}

We demonstrated two methods of shaping the~autocorrelation of optical intensity. The~first allows for more arbitrary shapes; the~second is limited to nonincreasing convex functions, but allows tailoring the~intensity and photon statistics as well.

Primarily, we consider our contribution to address a~fundamental problem in optical signal generation. However, possible practical uses involve exciting or probing systems sensitive to photon correlations \cite{Kazimierczuk2015Jul,Spasibko2017Nov}. The~modulation techniques may also serve to simulate optical fading channels and atmospheric scintillation \cite{Vasylyev2016Aug}, especially if certain power spectra are required \cite{Dravins1997,Toyoshima2011Aug,Shen2014}. Atmospheric turbulence has a~critical impact on both satellite-based \cite{Liao2018Jan,Yin2020Jun} and ground-to-ground \cite{Usenko2012Sep,Gong2018Jul,Ruppert2019Dec,Cao2020Dec} quantum communications. Especially, single-mode coupling of atmospheric channels results in significant transmission and phase fluctuations. The~proposed methods allow simulation of the~channel transmission. The~specific properties that can be reproduced are the~fading distribution and temporal correlations.

\begin{acknowledgments}
This research has been supported by QuantERA ERA-NET Cofund in Quantum Technologies, EU Horizon 2020, and Ministry of Education, Youth and Sports of the~Czech Republic (Project HYPER-U-P-S, No.~8C18002). M.~J. also acknowledges the~support of the~Czech Science Foundation (Grant No.~19-19189S).
\end{acknowledgments}

\appendix

\section{Derivation of autocorrelation of stochastic hill superposition}
\label{ap:derivation_hills}

Let the~``hill'' function be $h(t)$ and its unnormalized cross-correlation function
\begin{equation}
C(\tau) \coloneqq (h \star h)(\tau) = \int_{-\infty}^\infty h(t)h(t+\tau)\mathrm{d}t.
\end{equation}
We are going to consider the~intensity of light as a~sum of hills randomly distributed in time:
\begin{equation}
I(t)=\sum_i h(t-t_i),
\end{equation}
where the~peaks $t_i$ follow a~homogeneous Poisson point process in time with a~mean frequency $\lambda$. This means that peak-to-peak distances $\Delta t_i = t_{i+1}-t_i$ follow a~negative-exponential probability density distribution $p(\Delta t_i)=\lambda \exp(-\lambda\,\Delta t_i)$. The~normalized intensity autocorrelation function is
\begin{equation}
g^{(2)}(\tau) \coloneqq \frac{\langle I(t)I(t+\tau) \rangle}{\langle I(t) \rangle \langle I(t+\tau) \rangle}.
\end{equation}

Let us consider that the~intensity signal was generated in a~time span $T$ and assume the~limit of long $T$ in our calculations. Then, the~number of hills is $N \simeq \lambda\,T$. We can also denote the~integral
\begin{equation}
\|h\| \coloneqq \int_{-\infty}^\infty h(t)\mathrm{d}t,
\end{equation}
which also means $\int_{-\infty}^\infty C(\tau)\mathrm{d}\tau = \|h\|^2$. The~averaging terms can be calculated simply using the~linearity of integration,
\begin{align}\nonumber
\langle I(t) \rangle &= \langle I(t+\tau) \rangle \\
& = \frac{1}{T}\int_T I(t) \mathrm{d}t \simeq \frac{N \|h\|}{T}, \\\nonumber
\langle I(t)I(t+\tau) \rangle &= \frac{1}{T} \int_T \left(\sum_{i=1}^N h(t-t_i) \right) \\
 & \quad \times \left(  \sum_{j=1}^N h(t-t_j + \tau)\right) \\\nonumber
 &\simeq \frac{1}{T} \sum_{i,j} C(\Delta t_{ij} + \tau),
\end{align}
where $\Delta t_{ij}=t_j - t_i$ are distances between any two peaks. The~sum has $N^2$ terms that can be reduced by considering $\Delta t_{i=j} = 0$ and $\Delta t_{ij} = -\Delta t_{ji}$:

\begin{align} \label{eq.ACsplit}
\frac{1}{T} \sum_{i,j} C(\Delta t_{ij} + \tau) &= \frac{1}{T}\sum_{i<j} \Big[ C(\tau+\Delta t_{ij})  \\\nonumber
& \quad + C(\tau-\Delta t_{ij}) \Big] + \frac{N}{T} C(\tau).
\end{align}

The~peak locations $t_i$ are determined by sampling a~random Poisson process and we sum over a~large set of samples $\Delta t_{ij}$. We can therefore approximate the~above sum by

\begin{equation} \label{eq.ACsum}
\sum_{j=2}^N \sum_{i=1}^{j-1} C(\tau \pm \Delta t_{ij}) \simeq \frac{N^2-N}{2} \left\langle  C(\tau \pm \Delta t) \right\rangle_{\Delta t},
\end{equation}
where the~averaging follows a~certain probability distribution of distances $\Delta t$ between any two different peaks. Let us now calculate this distribution. By virtue of the~homogeneous Poisson process, we may consider any two different points in time independent with respect to the~random occurrence of samples. This allows us to define the~average density of peaks per time, which is constant: $\rho(0<t<T) = \lambda$. Then, the~density of peak pairs per distance $\Delta t$ is
\begin{equation}
\rho_2(\Delta t) = \int_0^{T-\Delta t} \rho(t) \rho(t+\Delta t) \mathrm{d}t = \lambda^2 (T-\Delta t),
\end{equation}
while the~corresponding probability density distribution follows by normalizing
\begin{equation} \label{eq.triangularDist}
p(\Delta t) = \frac{\rho_2(\Delta t)}{\int_0^T \rho_2(\Delta t)\mathrm{d}\Delta t} = \frac{2}{T}-\frac{2}{T^2} \Delta t.
\end{equation}

Now we can make use of this distribution to calculate \eqref{eq.ACsum}. Because we assume that time $T$ is much longer than both the~width of $C$ and delay $\tau$, we can either neglect the~linear term in \eqref{eq.triangularDist} or integrate by parts and approximate the~primitive function of $C$ by a~step function. In any way, we get

\begin{equation} \label{eq.ACmean}
\left\langle  C(\tau + \Delta t) + C(\tau - \Delta t) \right\rangle_{\Delta t} \simeq \frac{2\,\|h\|^2}{T}.
\end{equation}

Now we can consecutively substitute \eqref{eq.ACmean}, \eqref{eq.ACsum}, (\ref{eq.ACsplit}), use $N^2 \gg N$, and get
\begin{equation}
\langle I(t)I(t+\tau) \rangle \simeq \lambda C(\tau) + \frac{N^2 \|h\|^2}{T^2}.
\end{equation}

Finally, after substituting the~rest, we get the~final result for the~normalized autocorrelation of the~randomly generated signal
\begin{equation} \label{eq.g2}
g^{(2)}(\tau) = 1 + \frac{1}{\lambda} C_{\mathrm{norm}}(\tau),
\end{equation}
where $C_{\mathrm{norm}}(\tau) = C(\tau)/\|h\|^2$ is the~normalized cross-correlation of $h(t)$. We can see that the~shape of $g^{(2)}(\tau)$ does not depend on the~amplitude of $h(t)$, only on its shape. The~scaling of the~$g^{(2)}(\tau)$ is determined solely by the~mean frequency $\lambda$.

Factoring in the~background $I_\mathrm{bg}$ is a~straightforward extension to the~calculations above, yielding \eqref{eq.g2bg}. After maximizing with respect to $\lambda$, we get the~maximum bunching
\begin{equation}\label{eq.g2max}
g^{(2)}_\mathrm{max}(0) = 1 + \frac{\|h\|}{4I_\mathrm{bg}}C_{\mathrm{norm}}(0).
\end{equation}
Let us formulate the~dynamic range of the~modulation as
\begin{equation}
R_{\mathrm{D}} = h_\mathrm{max}/I_\mathrm{bg},
\end{equation}
which makes sense in the~limit of high extinction $R_{\mathrm{D}} \gg 1$ and $\lambda \gg 1$, where the~hills are far apart and peak value $h_\mathrm{max}$ of the~hill function effectively represents the~maximum intensity that needs to be generated. Then, we can rewrite \eqref{eq.g2max} as
\begin{equation}\label{eq.g2max2}
g^{(2)}_\mathrm{max}(0) = 1 + R_{\mathrm{D}}\frac{\|h\|}{4h_\mathrm{max}}C_{\mathrm{norm}}(0).
\end{equation}

The~shape of the~autocorrelation is given by $C_{\mathrm{norm}}(\tau)$, which is given by $h(t)$. Let us define that two autocorrelations $g_1$ and $g_2$ have the~same shape if and only if they are identical up to scaling factors $a$ and $b$,
\begin{equation}
g_1(\tau)-1 = a \left(g_2(b\tau)-1 \right).
\end{equation}
When considering the maximum vertical scale \eqref{eq.g2max2}, it can be shown that it is invariant under horizontal scaling while keeping the~dynamic range constant. One can easily check this by substituting $h(t) \to h(bt)$ into \eqref{eq.hnormdef}, \eqref{eq.Cdef}, and \eqref{eq.g2max2}. Thus we can conclude that the~maximum bunching is limited by the~modulation dynamic range and by the~shape of the~autocorrelation. Some examples are given in Table~\ref{tab3}.
\begin{table}[h]
\begin{ruledtabular}
\begin{tabular}{rl}
Gauss & $g^{(2)}_\mathrm{max}(0) = 1 + R_{\mathrm{D}}/\big( 4\sqrt{2} \big)$	\\
Cauchy--Lorentz & $g^{(2)}_\mathrm{max}(0) = 1 + R_{\mathrm{D}}/8$	\\
Hyperb.\@ secant & $g^{(2)}_\mathrm{max}(0) = 1 + R_{\mathrm{D}}/(2\pi)$	\\
Triangle & $g^{(2)}_\mathrm{max}(0) = 1 + R_{\mathrm{D}}/4$	\\
\end{tabular}
\end{ruledtabular}
\caption{Autocorrelation shapes and the~corresponding maximum bunching values.}
\label{tab3}
\end{table}

\section{Obtaining the~hill shape from autocorrelation}
\label{ap:hill_inverse}

Let us have a~given autocorrelation function by specifying \eqref{eq.g2} numerically or analytically. Our task is to determine $h(t)$ and $\lambda$, which together define the~desired signal. We use the~property of the~Fourier transform of a~cross-correlation, $\mathcal{F}(C) = |\mathcal{F}(h)|^2$. Therefore, we can obtain
\begin{equation} \label{eq.fourierInversion}
h(t) = \mathcal{F}^{-1}\left[\sign(\nu)\sqrt{\mathcal{F}[C(t)]}\right],
\end{equation}
where the~sign function is a~piecewise $\pm 1$. Because the~hill scale $\|h\|$ is arbitrary, we can simply assign $C(t) \leftarrow \left(g^{(2)}(t)-1\right)$ and scale the~resulting $h(t)$.

The~existence of a~precise solution depends on two conditions. The~Fourier transform of the~cross-correlation $C$ and the~result $h(t)$ must both be non-negative. The~non-negativity of $\mathcal{F}[C]$ is not a~restrictive condition, because any legal $g^{(2)}(\tau)$ that can arise from \eqref{eq:g2def} automatically complies due to the~Wiener-Khinchin theorem. The~positivity of the~resulting $h(t)$ can sometimes be ensured using a~suitable sign-function---a good example is a~triangular $C$ resulting in a~rectangular $h(t)$.

If the~autocorrelation is given numerically, we can employ discrete Fourier transform (DFT) defined by relations
\begin{align}
\text{DFT:}\quad X_k &= \sum_{n=0}^{N-1} x_n \mathrm{e}^{-i 2\pi n k / N},\\
\text{DFT$^{-1}$:}\quad x_n &= \frac{1}{N}\sum_{k=0}^{N-1} X_k \mathrm{e}^{i 2\pi nk/N}.
\end{align}

Let us consider a~given discrete series $\{c_n\} = c_0,c_1,\dots,c_{N-1}$ that defines the~autocorrelation symmetrically around zero with a~sampling interval $\delta t$,
\begin{equation}
c_n \coloneqq C\left[ \left(n-\frac{N-1}{2} \right) \delta t \right].
\end{equation}
Then, we can approximate the~Fourier transform
\begin{align}\nonumber
\widehat{C}(\nu) &= \int C(t) \exp(-i 2\pi t \nu) \mathrm{d}t \\
& \simeq \sum_n C(t_n) \exp(-i 2\pi t_n \nu) \delta t, \\\nonumber
\widehat{c}_k &= \widehat{C}\left( \frac{k}{N \delta t} \right) \frac{1}{\delta t} \\
& = \mathrm{DFT}(c_n) \exp \left( -i \pi \frac{N-1}{N} k \right).
\end{align}

This way, we obtain the~discrete Fourier transform with a~sampling period $\delta \nu = 1/(N \delta t)$. We proceed and get a~discrete version of \eqref{eq.fourierInversion},
\begin{equation}
h_m = \mathrm{DFT}^{-1}\left[(\pm 1)_k\sqrt{\widehat{c}_k}\right]\frac{1}{\sqrt{\delta t}}.
\end{equation}

If we consider the~periodicity of DFT, we know that $h_m = h_{m - N}$ and we can map the~implicit range $[0, N-1]$ to $[ -(N-1)/2,(N-1)/2 ]$ and obtain the~hill function
\begin{equation}
h(m \times \delta t) = 
\begin{cases}
h_m & 0 \leq m \leq \frac{N-1}{2}, \\
h_{m+N} & -\frac{N-1}{2} \leq m < 0,
\end{cases}
\quad m \in \mathds{Z}.
\end{equation}

\section{Derivation of nonoverlapping hills}
\label{ap:non-overlapping}

For high bunching, when the~mean delay $\langle \Delta t \rangle = 1/\lambda$ is much greater than the~width of the~autocorrelation, the~signal approaches a~sequence of randomly distributed, but disjoint pulses $h(t)$. If a~pulse generator is used, it cannot typically generate overlapping signals, which creates a~distortion in the~$g^{(2)}(\tau)$. If $h(t)$ is defined for $t \in (-t_0/2,t_0/2)$, the~minimum delay $\Delta t$ can be $t_0$. This means that the~peaks no longer follow a~Poisson process, but a~self-excited point process. The~limitation of $t_0$ is akin to nonparalyzable dead time of particle detectors.

The~distribution of peak pairs becomes more complicated. The~term in \eqref{eq.ACsplit} that sums all the~pairs needs to be evaluated separately for immediate neighbors, next-but-one neighbors, and so on. This is because the~distribution $p_k(\Delta t)$ of $\Delta t_{i,i+k}$ must be calculated for each $k=1,2,3,\dots$,

\begin{align}\label{eq.ACsumDT}
\sum_{i<j} C(\tau \pm \Delta t_{ij}) &= \sum_{k=1}^{N-1} \sum_{i=1}^{N-k} C(\tau \pm \Delta t_{i,i+k}) \\\nonumber
& \simeq \sum_{k=1}^{N-1} (N-k) \langle C(\tau \pm \Delta t) \rangle_{\Delta t \sim p_k}.
\end{align}

The~first step is to consider the~sequential delays $\Delta t_{i,i+1}$. By definition, these \emph{are} independent realizations of a~random Poisson process with an~addition of $t_0$. Therefore the~probability density function is a~shifted negative exponential
\begin{equation}
p_1(\Delta t) =
\begin{cases}
0 & \Delta t < t_0	\\
\lambda \mathrm{e}^{-\lambda (\Delta t - t_0)} & \Delta t \geq t_0
\end{cases}.
\end{equation}
Now, considering that $\Delta t_{i,i+2} = \Delta t_{i,i+1} + \Delta t_{i+1,i+2}$, the~constituent delays are mutually independent with known distributions. That means that the~result has a~known distribution. By further induction, as more independent terms are added, we can average over them in each step,
\begin{align}\nonumber
p_k(\Delta t) &= \int_{t'=0}^{\Delta t} p_{k-1}(t') p_1(\Delta t - t') \mathrm{d}t' \\
& =
\begin{cases}
0 & \Delta t < k t_0	\\
\lambda^k \frac{( \Delta t - k t_0 )^{k-1}}{(k-1)!} \mathrm{e}^{-\lambda (\Delta t - k t_0)} & \Delta t \geq k t_0,
\end{cases}
\end{align}
where $\Delta t$ in $p_k$ means $\Delta t_{i,i+k}$, describing its distribution over all $i$ for $N \to \infty$.

As we assume the~limit of long signals, but are interested in finite delays, let us fix $\tau$ and calculate the~limit $N \to \infty$. After substitution of $p_k$ into \eqref{eq.ACsumDT}, the~averaged terms become
\begin{align}\nonumber
\langle C(\tau \pm \Delta t)\rangle_{p_k} & = \int_{t=0}^\infty C(\tau \pm t \pm k t_0) \\
\label{eq.ACmeanDT}
& \qquad \times \lambda^k \frac{t^{k-1}}{(k-1)!}\mathrm{e}^{-\lambda t} \mathrm{d}t.
\end{align}
Now let us recall that $C(t)$ has a~finite support and so for indices $k > |\tau|/t_0 + 1$, the~first term $C(\tau \pm t \pm k t_0) \equiv 0$ during the~whole integration for both plus and minus signs. This means that for each finite $\tau$, the~number of nonzero terms in \eqref{eq.ACsumDT} is finite. As a~result, $k/N \to 0$ for each $k$. Also note that $T/N \to (1/\lambda + t_0)$. The~result then is
\begin{align}\nonumber
g^{(2)}(\tau) & = \left( \frac{1}{\lambda} + t_0\right) \sum_{k=1}^{\lfloor |\tau|/t_0 + 1 \rfloor} \int_{t=0}^{-(k-1)t_0 + |\tau|} \\\nonumber
& \qquad \left[ C_{\mathrm{norm}}(\tau + t + kt_0) + C_{\mathrm{norm}}(\tau - t - kt_0) \right] \\\nonumber
& \qquad \times \lambda^k \frac{t^{k-1}}{(k-1)!}\mathrm{e}^{-\lambda t} \mathrm{d}t	\\
& \quad+ \left( \frac{1}{\lambda} + t_0\right)C_{\mathrm{norm}}(\tau).
\end{align}
This result also trivially holds for the~case when the~minimum delay is $t_0$, but the~support of $h(t)$ is narrower. Moreover, the~formula can also be adopted for the~case of infinite support of $h(t)$ and $C(t)$ with a~minimum delay $t_0$. In that case, both the~sum and the~integral would go to infinity; otherwise, the~formula would be the~same. The~only difference in the~derivation would be in proving the~limit in \eqref{eq.ACsumDT},
\begin{equation}
\lim_{N \to \infty} \sum_{k=1}^{N-1} \frac{k}{N} \langle C(\tau \pm \Delta t) \rangle_k = 0.
\end{equation}
As $C(t)$ must be integrable, its tail scales faster than $1/t$. From \eqref{eq.ACmeanDT} and the~integrability of the~gammalike kernel, it follows that from a~certain $k>k_0$, the~term $\langle C(\tau \pm \Delta t) \rangle_k$ decreases faster than $1/k$, which means that the~limit of the~series is zero.

\section{Derivation of random switching of intensity levels}
\label{ap:derivation_levels}

The~signal $I(t)$ is now defined as a~step function, where each intensity level $I_i$ is held for a~time period $\Delta t_i$. Each of these variables is independently governed by its probability density $p(I)$, $p(\Delta t)$. Let us denote averaging over these distributions by unmarked angle brackets $\langle \cdot \rangle$, where the~distribution is determined by the~random variable inside. Averaging over time will be marked by $\langle \cdot \rangle_t$. If we denote the~overall number of steps as $N$ and the~overall time as $T$, we get the~mean intensity,
\begin{align}\nonumber
\langle I(t) \rangle_t &= \frac{1}{T} \sum_{i=1}^N \Delta t_i I_i = \frac{N}{T} \langle \Delta t I \rangle_t \\
&= \frac{1}{\langle \Delta t \rangle} \langle \Delta t \rangle \langle I \rangle = \langle I \rangle.
\end{align}
This result easily follows from the~uncorrelation of $\Delta t$ and $I$. However, more discussion is needed for the~autocorrelation term.

We express the~integral as a~sum over the~steps again, but the~term depends on the~delay $\tau$ being greater or smaller than the~step width $\Delta t_i$,
\begin{multline}\label{eq:overlap1}
\int_0^T I(t) I(t+\tau) \mathrm{d}t \\
= \sum_i
\begin{cases}
(\Delta t_i - \tau) I_i^2 + \tau I_i I_i^\mathrm{eff} & \Delta t_i > \tau \\
\Delta t_i I_i I_i^\mathrm{eff} & \Delta t_i \leq \tau
\end{cases}
.
\end{multline}

\begin{figure}
\centering
\includegraphics[width=.8\linewidth]{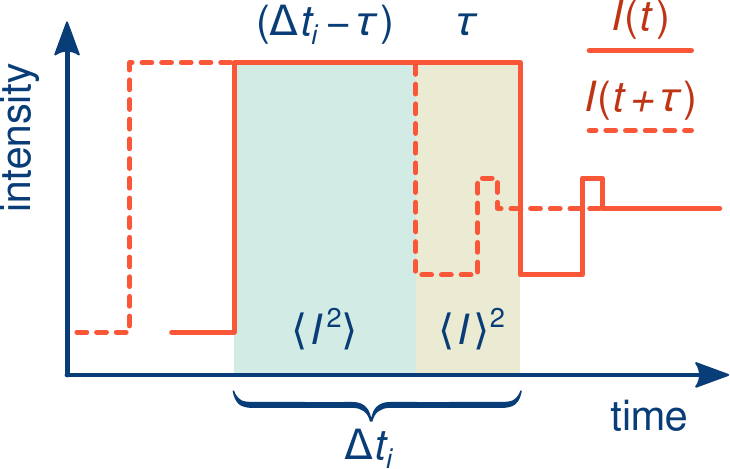}
\caption{The~illustration of the~integral overlap. The~highlighted area marks the~integration over the~$i$-th step. After averaging, the~two integral parts are revealed to correspond to the~terms $\langle I^2 \rangle$ and $\langle I \rangle^2$.}
\label{fig:level_overlap}
\end{figure}

The~case of $\Delta t_i > \tau$ is illustrated in Fig.~\ref{fig:level_overlap}. The~$\langle I^2 \rangle$ area on the~left corresponds to the~overlap where both signals have the~intensity $I_i$. The~$\langle I \rangle^2$ area on the~right is a~product of $I(t) = I_i$ and the~random intensity levels in $I(t+\tau)$ that follow the~$i$th level. We can express this integral overlap as $\tau I_i I_i^\mathrm{eff}$, where $I_i^\mathrm{eff}$ is the~\emph{average} intensity of $I(t+\tau)$ in the~area of width $\tau$. If $\Delta t_i \leq \tau$, the~$i$th step has zero overlap with itself and only one term remains.

Now, if we consider the~summation over $i$, we may again replace it with averaging over $\Delta t$ and $I$ independently. What is more, the~variable $I_i^\mathrm{eff}$ gets averaged independently of $I_i$, because the~intensity levels are mutually uncorrelated as well. Therefore $I_i^\mathrm{eff} \to \langle I \rangle$, $I_i \to \langle I \rangle$, and $I_i^2 \to \langle I^2 \rangle$. However, the~averaging over $\Delta t$ needs to be split due to the~piecewise definition in \eqref{eq:overlap1}. We obtain
\begin{multline}
\int_0^T I(t) I(t+\tau) \mathrm{d}t \\
= N \int_\tau^\infty p(\Delta t) \left[ (\Delta t - \tau)\langle I^2 \rangle + \tau \langle I \rangle^2 \right] \mathrm{d}\Delta t \\
+ N \int_0^\tau p(\Delta t) \Delta t \langle I \rangle^2 \mathrm{d}\Delta t.
\end{multline}
After extending the~second integral to infinity to obtain the~term $N \langle \Delta t \rangle \langle I \rangle^2$ and subtracting this extension from the~integral on the~left, we get
\begin{align}\nonumber
g^{(2)}(\tau) &= \frac{1}{T\langle I \rangle^2} \int_0^T I(t) I(t+\tau) \mathrm{d}t \\
&= 1 + \frac{\langle I^2 \rangle - \langle I \rangle^2}{\langle \Delta t \rangle \langle I \rangle^2} \int_\tau^\infty p(\Delta t)(\Delta t - \tau) \mathrm{d}\Delta t.
\end{align}

\section{Existence of a~solution for random intensity levels}
\label{ap:existence_levels}

The~necessary conditions are that the~given autocorrelation is continuous, bounded ($g^{(2)}(0) > 1$ and $\lim_{\tau \to \infty}g^{(2)}(\tau) = 1$), monotone, and convex. We claim that these conditions are also sufficient for the~existence of a~solution. We are going to simplify the~notation of the~autocorrelation and its derivatives to $g,g',g''$.

First, we propose that
\begin{equation}\label{eq:g'limit}
\lim_{\tau \to \infty} g'(\tau) = 0.
\end{equation}
Because we know that $g'$ is monotone (convexity of $g$) and upper bounded ($g'(\tau) \leq 0$), there must be a~limit due to the~monotone convergence theorem. We can prove by contradiction that the~limit is zero.

Let us assume that $\lim_{\tau \to \infty} g'(\tau) = L$, where $L < 0$. Then there exists $\tau_0$ such that $\forall \tau \geq \tau_0: g'(\tau) \leq L/2$. We can formulate the~inequality
\begin{align}
\int_0^\infty g'(\tau)\mathrm{d}\tau &= \int_0^{\tau_0} g'(\tau)\mathrm{d}\tau + \int_{\tau_0}^\infty g'(\tau)\mathrm{d}\tau \\
&\leq \int_0^{\tau_0} g'(\tau)\mathrm{d}\tau + \int_{\tau_0}^\infty \frac{L}{2}\mathrm{d}\tau = -\infty.
\end{align}
This divergence is in conflict with the~existence of a~converging primitive function $g$, where $\int_0^\infty g'(\tau) = 1-g(0)$. Because $L$ obviously cannot be positive ($g'(\tau) \leq 0$), $L=0$ and the~limit \eqref{eq:g'limit} is proven.

Using Eqs.~\eqref{eq:g2_derivative1} and \eqref{eq:g2_derivative2}, we see that the~probability density can be expressed as
\begin{equation}
p(\Delta t) = -\frac{g''(\Delta t)}{g'(0)}.
\end{equation}
The~initial conditions ($g' \leq 0, g'' \geq 0$) mean that $p(\Delta t)$ is non-negative and owing to \eqref{eq:g'limit}, it is also normalized,
\begin{equation}
\int_0^\infty p(\Delta t)\mathrm{d}\Delta t = 1.
\end{equation}

The~only assumption that is left to prove is that the~mean value $\langle \Delta t \rangle$ is finite. To this end, we need to establish a~survival function as an~alternative to the~probability density,
\begin{equation}\label{eq:survival_formula}
S(t) \coloneqq \Pr[\Delta t > t] = \int_t^\infty p(\Delta t)\mathrm{d}\Delta t = \frac{g'(t)}{g'(0)}.
\end{equation}
A fundamental property of the~survival function is
\begin{equation}\label{eq:survival_derivative}
\frac{\mathrm{d}S(t)}{\mathrm{d}t} = -p(t),
\end{equation}
as well as its image $S(t) \in (0,1]$. The~image is satisfied for all $t \geq 0$ owing to \eqref{eq:survival_formula} and \eqref{eq:g'limit}.

The~standard formula for the~mean value is
\begin{equation}\label{eq:mean_delay}
\langle \Delta t \rangle \coloneqq \int_0^\infty \Delta t \times p(\Delta t)\mathrm{d}\Delta t.
\end{equation}
However, the~survival function offers an~alternative \cite{Lo2018Sep},
\begin{equation}
\langle \Delta t \rangle = \int_0^\infty S(t)\mathrm{d}t.
\end{equation}

\begin{figure}
	\centering
	\includegraphics[width=0.8\linewidth]{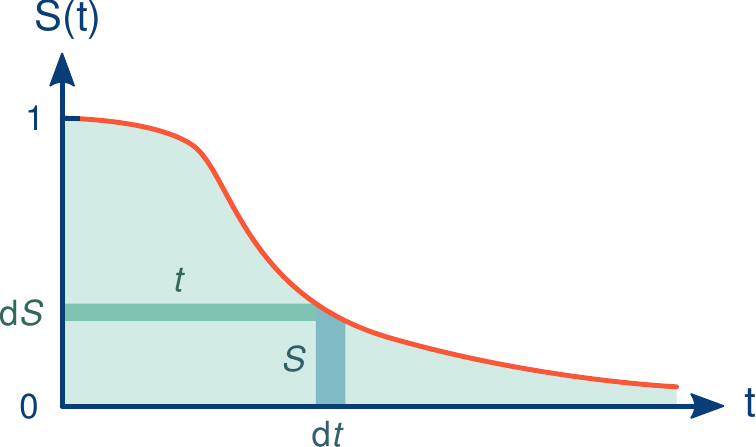}
	\caption{The~integral can be computed by summing horizontal or vertical slices.}
	\label{fig:survival}
\end{figure}

The~equivalence of these formulas is illustrated in Fig.~\ref{fig:survival} and proven by the~following:
\begin{align}
\int_0^\infty S(t)\mathrm{d}t &= \lim_{S_0 \to 0} \int_{S_0}^1 t(S)\mathrm{d}S = \lim_{t_0 \to \infty} \int_{t_0}^0 t \frac{\mathrm{d}S}{\mathrm{d}t} \mathrm{d}t \\
&= \lim_{t_0 \to \infty} \int_0^{t_0} t \times p(t) \mathrm{d}t
\end{align}
In the~first step, we switch to integration of horizontal slices $t\times\mathrm{d}S$. Then, we change the~integration variable $S \to t$, substitute \eqref{eq:survival_derivative}, and arrive at \eqref{eq:mean_delay}. The~integrability of $S(t)$ follows from \eqref{eq:survival_formula} and the~properties of $g(t)$. We therefore obtain a~finite mean value
\begin{equation}
\langle \Delta t \rangle = \frac{1-g(0)}{g'(0)}.
\end{equation}

In conclusion, we have proved the~existence of a~solution---a valid finite-mean probability distribution of $\Delta t$---for every monotone, convex, and appropriately bounded $g^{(2)}(\tau)$.

\bibliography{g2references}

\begin{thebibliography}{32}%
\makeatletter
\providecommand \@ifxundefined [1]{%
 \@ifx{#1\undefined}
}%
\providecommand \@ifnum [1]{%
 \ifnum #1\expandafter \@firstoftwo
 \else \expandafter \@secondoftwo
 \fi
}%
\providecommand \@ifx [1]{%
 \ifx #1\expandafter \@firstoftwo
 \else \expandafter \@secondoftwo
 \fi
}%
\providecommand \natexlab [1]{#1}%
\providecommand \enquote  [1]{``#1''}%
\providecommand \bibnamefont  [1]{#1}%
\providecommand \bibfnamefont [1]{#1}%
\providecommand \citenamefont [1]{#1}%
\providecommand \href@noop [0]{\@secondoftwo}%
\providecommand \href [0]{\begingroup \@sanitize@url \@href}%
\providecommand \@href[1]{\@@startlink{#1}\@@href}%
\providecommand \@@href[1]{\endgroup#1\@@endlink}%
\providecommand \@sanitize@url [0]{\catcode `\\12\catcode `\$12\catcode
  `\&12\catcode `\#12\catcode `\^12\catcode `\_12\catcode `\%12\relax}%
\providecommand \@@startlink[1]{}%
\providecommand \@@endlink[0]{}%
\providecommand \url  [0]{\begingroup\@sanitize@url \@url }%
\providecommand \@url [1]{\endgroup\@href {#1}{\urlprefix }}%
\providecommand \urlprefix  [0]{URL }%
\providecommand \Eprint [0]{\href }%
\providecommand \doibase [0]{https://doi.org/}%
\providecommand \selectlanguage [0]{\@gobble}%
\providecommand \bibinfo  [0]{\@secondoftwo}%
\providecommand \bibfield  [0]{\@secondoftwo}%
\providecommand \translation [1]{[#1]}%
\providecommand \BibitemOpen [0]{}%
\providecommand \bibitemStop [0]{}%
\providecommand \bibitemNoStop [0]{.\EOS\space}%
\providecommand \EOS [0]{\spacefactor3000\relax}%
\providecommand \BibitemShut  [1]{\csname bibitem#1\endcsname}%
\let\auto@bib@innerbib\@empty
\bibitem [{\citenamefont {Loudon}(2000)}]{Loudon}%
  \BibitemOpen
  \bibfield  {author} {\bibinfo {author} {\bibfnamefont {R.}~\bibnamefont
  {Loudon}},\ }\href@noop {} {\emph {\bibinfo {title} {The Quantum Theory of
  Light}}},\ \bibinfo {edition} {3rd}\ ed.\ (\bibinfo  {publisher} {Oxford
  University Press},\ \bibinfo {address} {Oxford, U.K.},\ \bibinfo {year}
  {2000})\BibitemShut {NoStop}%
\bibitem [{\citenamefont {Martienssen}\ and\ \citenamefont
  {Spiller}(2005)}]{Martienssen2005Jul}%
  \BibitemOpen
  \bibfield  {author} {\bibinfo {author} {\bibfnamefont {W.}~\bibnamefont
  {Martienssen}}\ and\ \bibinfo {author} {\bibfnamefont {E.}~\bibnamefont
  {Spiller}},\ }\bibfield  {title} {\bibinfo {title} {{Coherence and
  Fluctuations in Light Beams}},\ }\href {https://doi.org/10.1119/1.1970023}
  {\bibfield  {journal} {\bibinfo  {journal} {American Journal of Physics}\
  }\textbf {\bibinfo {volume} {32}},\ \bibinfo {pages} {919} (\bibinfo {year}
  {2005})}\BibitemShut {NoStop}%
\bibitem [{\citenamefont {Boitier}\ \emph {et~al.}(2009)\citenamefont
  {Boitier}, \citenamefont {Godard}, \citenamefont {Rosencher},\ and\
  \citenamefont {Fabre}}]{Boitier2009Apr}%
  \BibitemOpen
  \bibfield  {author} {\bibinfo {author} {\bibfnamefont {F.}~\bibnamefont
  {Boitier}}, \bibinfo {author} {\bibfnamefont {A.}~\bibnamefont {Godard}},
  \bibinfo {author} {\bibfnamefont {E.}~\bibnamefont {Rosencher}},\ and\
  \bibinfo {author} {\bibfnamefont {C.}~\bibnamefont {Fabre}},\ }\bibfield
  {title} {\bibinfo {title} {{Measuring photon bunching at ultrashort timescale
  by two-photon absorption in semiconductors}},\ }\href
  {https://doi.org/10.1038/nphys1218} {\bibfield  {journal} {\bibinfo
  {journal} {Nature Physics}\ }\textbf {\bibinfo {volume} {5}},\ \bibinfo
  {pages} {267} (\bibinfo {year} {2009})}\BibitemShut {NoStop}%
\bibitem [{\citenamefont {Nazarathy}\ \emph {et~al.}(1989)\citenamefont
  {Nazarathy}, \citenamefont {Newton}, \citenamefont {Giffard}, \citenamefont
  {Moberly}, \citenamefont {Sischka}, \citenamefont {Trutna},\ and\
  \citenamefont {Foster}}]{Nazarathy1989Jan}%
  \BibitemOpen
  \bibfield  {author} {\bibinfo {author} {\bibfnamefont {M.}~\bibnamefont
  {Nazarathy}}, \bibinfo {author} {\bibfnamefont {S.~A.}\ \bibnamefont
  {Newton}}, \bibinfo {author} {\bibfnamefont {R.~P.}\ \bibnamefont {Giffard}},
  \bibinfo {author} {\bibfnamefont {D.~S.}\ \bibnamefont {Moberly}}, \bibinfo
  {author} {\bibfnamefont {F.}~\bibnamefont {Sischka}}, \bibinfo {author}
  {\bibfnamefont {W.~R.}\ \bibnamefont {Trutna}},\ and\ \bibinfo {author}
  {\bibfnamefont {S.}~\bibnamefont {Foster}},\ }\bibfield  {title} {\bibinfo
  {title} {{Real-time long range complementary correlation optical time domain
  reflectometer}},\ }\href {https://doi.org/10.1109/50.17729} {\bibfield
  {journal} {\bibinfo  {journal} {Journal of Lightwave Technology}\ }\textbf
  {\bibinfo {volume} {7}},\ \bibinfo {pages} {24} (\bibinfo {year}
  {1989})}\BibitemShut {NoStop}%
\bibitem [{\citenamefont {Bennink}\ \emph {et~al.}(2002)\citenamefont
  {Bennink}, \citenamefont {Bentley},\ and\ \citenamefont
  {Boyd}}]{Bennink2002Aug}%
  \BibitemOpen
  \bibfield  {author} {\bibinfo {author} {\bibfnamefont {R.~S.}\ \bibnamefont
  {Bennink}}, \bibinfo {author} {\bibfnamefont {S.~J.}\ \bibnamefont
  {Bentley}},\ and\ \bibinfo {author} {\bibfnamefont {R.~W.}\ \bibnamefont
  {Boyd}},\ }\bibfield  {title} {\bibinfo {title}
  {{{\textasciigrave}{\textasciigrave}Two-Photon'' Coincidence Imaging with a
  Classical Source}},\ }\href {https://doi.org/10.1103/PhysRevLett.89.113601}
  {\bibfield  {journal} {\bibinfo  {journal} {Physical Review Letters}\
  }\textbf {\bibinfo {volume} {89}},\ \bibinfo {pages} {113601} (\bibinfo
  {year} {2002})}\BibitemShut {NoStop}%
\bibitem [{\citenamefont {Gatti}\ \emph {et~al.}(2004)\citenamefont {Gatti},
  \citenamefont {Brambilla}, \citenamefont {Bache},\ and\ \citenamefont
  {Lugiato}}]{Gatti2004Aug}%
  \BibitemOpen
  \bibfield  {author} {\bibinfo {author} {\bibfnamefont {A.}~\bibnamefont
  {Gatti}}, \bibinfo {author} {\bibfnamefont {E.}~\bibnamefont {Brambilla}},
  \bibinfo {author} {\bibfnamefont {M.}~\bibnamefont {Bache}},\ and\ \bibinfo
  {author} {\bibfnamefont {L.~A.}\ \bibnamefont {Lugiato}},\ }\bibfield
  {title} {\bibinfo {title} {{Ghost Imaging with Thermal Light: Comparing
  Entanglement and ClassicalCorrelation}},\ }\href
  {https://doi.org/10.1103/PhysRevLett.93.093602} {\bibfield  {journal}
  {\bibinfo  {journal} {Physical Review Letters}\ }\textbf {\bibinfo {volume}
  {93}},\ \bibinfo {pages} {093602} (\bibinfo {year} {2004})}\BibitemShut
  {NoStop}%
\bibitem [{\citenamefont {Zhou}\ \emph {et~al.}(2017)\citenamefont {Zhou},
  \citenamefont {Li}, \citenamefont {Bai}, \citenamefont {Chen}, \citenamefont
  {Liu}, \citenamefont {Xu},\ and\ \citenamefont {Zheng}}]{Zhou2017May}%
  \BibitemOpen
  \bibfield  {author} {\bibinfo {author} {\bibfnamefont {Y.}~\bibnamefont
  {Zhou}}, \bibinfo {author} {\bibfnamefont {F.-l.}\ \bibnamefont {Li}},
  \bibinfo {author} {\bibfnamefont {B.}~\bibnamefont {Bai}}, \bibinfo {author}
  {\bibfnamefont {H.}~\bibnamefont {Chen}}, \bibinfo {author} {\bibfnamefont
  {J.}~\bibnamefont {Liu}}, \bibinfo {author} {\bibfnamefont {Z.}~\bibnamefont
  {Xu}},\ and\ \bibinfo {author} {\bibfnamefont {H.}~\bibnamefont {Zheng}},\
  }\bibfield  {title} {\bibinfo {title} {{Superbunching pseudothermal light}},\
  }\href {https://doi.org/10.1103/PhysRevA.95.053809} {\bibfield  {journal}
  {\bibinfo  {journal} {Physical Review A}\ }\textbf {\bibinfo {volume} {95}},\
  \bibinfo {pages} {053809} (\bibinfo {year} {2017})}\BibitemShut {NoStop}%
\bibitem [{\citenamefont {Jechow}\ \emph {et~al.}(2013)\citenamefont {Jechow},
  \citenamefont {Seefeldt}, \citenamefont {Kurzke}, \citenamefont {Heuer},\
  and\ \citenamefont {Menzel}}]{Jechow2013Dec}%
  \BibitemOpen
  \bibfield  {author} {\bibinfo {author} {\bibfnamefont {A.}~\bibnamefont
  {Jechow}}, \bibinfo {author} {\bibfnamefont {M.}~\bibnamefont {Seefeldt}},
  \bibinfo {author} {\bibfnamefont {H.}~\bibnamefont {Kurzke}}, \bibinfo
  {author} {\bibfnamefont {A.}~\bibnamefont {Heuer}},\ and\ \bibinfo {author}
  {\bibfnamefont {R.}~\bibnamefont {Menzel}},\ }\bibfield  {title} {\bibinfo
  {title} {{Enhanced two-photon excited fluorescence from imaging agents using
  true thermal light}},\ }\href {https://doi.org/10.1038/nphoton.2013.271}
  {\bibfield  {journal} {\bibinfo  {journal} {Nature Photonics}\ }\textbf
  {\bibinfo {volume} {7}},\ \bibinfo {pages} {973} (\bibinfo {year}
  {2013})}\BibitemShut {NoStop}%
\bibitem [{\citenamefont {Wang}\ \emph {et~al.}(2008)\citenamefont {Wang},
  \citenamefont {Wang},\ and\ \citenamefont {Wang}}]{Wang2008Jul}%
  \BibitemOpen
  \bibfield  {author} {\bibinfo {author} {\bibfnamefont {Y.}~\bibnamefont
  {Wang}}, \bibinfo {author} {\bibfnamefont {B.}~\bibnamefont {Wang}},\ and\
  \bibinfo {author} {\bibfnamefont {A.}~\bibnamefont {Wang}},\ }\bibfield
  {title} {\bibinfo {title} {{Chaotic Correlation Optical Time Domain
  Reflectometer Utilizing Laser Diode}},\ }\href
  {https://doi.org/10.1109/LPT.2008.2002745} {\bibfield  {journal} {\bibinfo
  {journal} {IEEE Photonics Technology Letters}\ }\textbf {\bibinfo {volume}
  {20}},\ \bibinfo {pages} {1636} (\bibinfo {year} {2008})}\BibitemShut
  {NoStop}%
\bibitem [{\citenamefont {Lin}\ and\ \citenamefont {Liu}(2004)}]{Lin2004Dec}%
  \BibitemOpen
  \bibfield  {author} {\bibinfo {author} {\bibfnamefont {F.-Y.}\ \bibnamefont
  {Lin}}\ and\ \bibinfo {author} {\bibfnamefont {J.-M.}\ \bibnamefont {Liu}},\
  }\bibfield  {title} {\bibinfo {title} {{Chaotic lidar}},\ }\href
  {https://doi.org/10.1109/JSTQE.2004.835296} {\bibfield  {journal} {\bibinfo
  {journal} {IEEE Journal of Selected Topics in Quantum Electronics}\ }\textbf
  {\bibinfo {volume} {10}},\ \bibinfo {pages} {991} (\bibinfo {year}
  {2004})}\BibitemShut {NoStop}%
\bibitem [{\citenamefont {Kim}\ \emph {et~al.}(2010)\citenamefont {Kim},
  \citenamefont {Huh},\ and\ \citenamefont {Suh}}]{Kim2010}%
  \BibitemOpen
  \bibfield  {author} {\bibinfo {author} {\bibfnamefont {J.~S.}\ \bibnamefont
  {Kim}}, \bibinfo {author} {\bibfnamefont {Y.}~\bibnamefont {Huh}},\ and\
  \bibinfo {author} {\bibfnamefont {M.~W.}\ \bibnamefont {Suh}},\ }\bibfield
  {title} {\bibinfo {title} {A method to generate autocorrelated stochastic
  signals based on the random phase spectrum},\ }\href
  {https://doi.org/10.1080/14685240802528443} {\bibfield  {journal} {\bibinfo
  {journal} {Journal of the Textile Institute}\ }\textbf {\bibinfo {volume}
  {101}},\ \bibinfo {pages} {471} (\bibinfo {year} {2010})}\BibitemShut
  {NoStop}%
\bibitem [{\citenamefont {Müller}\ \emph {et~al.}(2012)\citenamefont
  {Müller}, \citenamefont {Rojas},\ and\ \citenamefont
  {Goodwin}}]{Mueller2012}%
  \BibitemOpen
  \bibfield  {author} {\bibinfo {author} {\bibfnamefont {C.}~\bibnamefont
  {Müller}}, \bibinfo {author} {\bibfnamefont {C.~R.}\ \bibnamefont {Rojas}},\
  and\ \bibinfo {author} {\bibfnamefont {G.~C.}\ \bibnamefont {Goodwin}},\
  }\bibfield  {title} {\bibinfo {title} {Generation of amplitude constrained
  signals with a prescribed spectrum},\ }\href
  {https://doi.org/10.1016/j.automatica.2011.09.038} {\bibfield  {journal}
  {\bibinfo  {journal} {Automatica}\ }\textbf {\bibinfo {volume} {48}},\
  \bibinfo {pages} {153} (\bibinfo {year} {2012})}\BibitemShut {NoStop}%
\bibitem [{\citenamefont {Liu}\ and\ \citenamefont {Munson}(1982)}]{Liu1982}%
  \BibitemOpen
  \bibfield  {author} {\bibinfo {author} {\bibfnamefont {B.}~\bibnamefont
  {Liu}}\ and\ \bibinfo {author} {\bibfnamefont {D.}~\bibnamefont {Munson}},\
  }\bibfield  {title} {\bibinfo {title} {Generation of a random sequence having
  a jointly specified marginal distribution and autocovariance},\ }\href
  {https://doi.org/10.1109/tassp.1982.1163990} {\bibfield  {journal} {\bibinfo
  {journal} {{IEEE} Transactions on Acoustics, Speech, and Signal Processing}\
  }\textbf {\bibinfo {volume} {30}},\ \bibinfo {pages} {973} (\bibinfo {year}
  {1982})}\BibitemShut {NoStop}%
\bibitem [{\citenamefont {Hunter}\ and\ \citenamefont
  {Kearney}(1983)}]{Hunter1983Jun}%
  \BibitemOpen
  \bibfield  {author} {\bibinfo {author} {\bibfnamefont {I.~W.}\ \bibnamefont
  {Hunter}}\ and\ \bibinfo {author} {\bibfnamefont {R.~E.}\ \bibnamefont
  {Kearney}},\ }\bibfield  {title} {\bibinfo {title} {{Generation of random
  sequences with jointly specified probability density and autocorrelation
  functions}},\ }\href {https://doi.org/10.1007/BF00337087} {\bibfield
  {journal} {\bibinfo  {journal} {Biological Cybernetics}\ }\textbf {\bibinfo
  {volume} {47}},\ \bibinfo {pages} {141} (\bibinfo {year} {1983})}\BibitemShut
  {NoStop}%
\bibitem [{\citenamefont {Leneman}(1967)}]{Leneman1967Sep}%
  \BibitemOpen
  \bibfield  {author} {\bibinfo {author} {\bibfnamefont {O.~A.~Z.}\
  \bibnamefont {Leneman}},\ }\bibfield  {title} {\bibinfo {title} {{Correlation
  Function and Power Spectrun of Randomly Shaped Pulse Trains}},\ }\href
  {https://doi.org/10.1109/TAES.1967.5408864} {\bibfield  {journal} {\bibinfo
  {journal} {IEEE Transactions on Aerospace and Electronic Systems}\ }\textbf
  {\bibinfo {volume} {AES-3}},\ \bibinfo {pages} {774} (\bibinfo {year}
  {1967})}\BibitemShut {NoStop}%
\bibitem [{\citenamefont {Pandey}\ \emph {et~al.}(2014)\citenamefont {Pandey},
  \citenamefont {Satapathy}, \citenamefont {Suryabrahmam}, \citenamefont
  {Ivan},\ and\ \citenamefont {Ramachandran}}]{Pandey2014Jun}%
  \BibitemOpen
  \bibfield  {author} {\bibinfo {author} {\bibfnamefont {D.}~\bibnamefont
  {Pandey}}, \bibinfo {author} {\bibfnamefont {N.}~\bibnamefont {Satapathy}},
  \bibinfo {author} {\bibfnamefont {B.}~\bibnamefont {Suryabrahmam}}, \bibinfo
  {author} {\bibfnamefont {J.~S.}\ \bibnamefont {Ivan}},\ and\ \bibinfo
  {author} {\bibfnamefont {H.}~\bibnamefont {Ramachandran}},\ }\bibfield
  {title} {\bibinfo {title} {{Classical light sources with tunable temporal
  coherence and tailored photon number distributions}},\ }\href
  {https://doi.org/10.1140/epjp/i2014-14115-2} {\bibfield  {journal} {\bibinfo
  {journal} {European Physical Journal Plus}\ }\textbf {\bibinfo {volume}
  {129}},\ \bibinfo {pages} {115} (\bibinfo {year} {2014})}\BibitemShut
  {NoStop}%
\bibitem [{\citenamefont {Straka}\ \emph {et~al.}(2018)\citenamefont {Straka},
  \citenamefont {Mika},\ and\ \citenamefont {Ježek}}]{Straka2018Apr}%
  \BibitemOpen
  \bibfield  {author} {\bibinfo {author} {\bibfnamefont {I.}~\bibnamefont
  {Straka}}, \bibinfo {author} {\bibfnamefont {J.}~\bibnamefont {Mika}},\ and\
  \bibinfo {author} {\bibfnamefont {M.}~\bibnamefont {Ježek}},\ }\bibfield
  {title} {\bibinfo {title} {Generator of arbitrary classical photon
  statistics},\ }\href {https://doi.org/10.1364/OE.26.008998} {\bibfield
  {journal} {\bibinfo  {journal} {Optics Express}\ }\textbf {\bibinfo {volume}
  {26}},\ \bibinfo {pages} {8998} (\bibinfo {year} {2018})}\BibitemShut
  {NoStop}%
\bibitem [{\citenamefont {Straka}\ and\ \citenamefont {Ježek}(2020)}]{Code}%
  \BibitemOpen
  \bibfield  {author} {\bibinfo {author} {\bibfnamefont {I.}~\bibnamefont
  {Straka}}\ and\ \bibinfo {author} {\bibfnamefont {M.}~\bibnamefont
  {Ježek}},\ }\href@noop {} {\bibinfo {title} {Shaping the $g^{(2)}$
  autocorrelation and photon statistics}},\ \bibinfo {howpublished} {Code
  Ocean} (\bibinfo {year} {2020}),\ \bibinfo {note}
  {\href{https://doi.org/10.24433/CO.4221408.v1}{doi:
  10.24433/CO.4221408.v1}}\BibitemShut {NoStop}%
\bibitem [{\citenamefont {Mandel}\ and\ \citenamefont
  {Wolf}(1995)}]{MandelWolf}%
  \BibitemOpen
  \bibfield  {author} {\bibinfo {author} {\bibfnamefont {L.}~\bibnamefont
  {Mandel}}\ and\ \bibinfo {author} {\bibfnamefont {E.}~\bibnamefont {Wolf}},\
  }\href {https://doi.org/10.1017/CBO9781139644105} {\emph {\bibinfo {title}
  {Optical Coherence and Quantum Optics}}}\ (\bibinfo  {publisher} {Cambridge
  University Press},\ \bibinfo {address} {Cambridge, U.K.},\ \bibinfo {year}
  {1995})\BibitemShut {NoStop}%
\bibitem [{\citenamefont {Kazimierczuk}\ \emph {et~al.}(2015)\citenamefont
  {Kazimierczuk}, \citenamefont {Schmutzler}, \citenamefont {A{\ss}mann},
  \citenamefont {Schneider}, \citenamefont {Kamp}, \citenamefont
  {H\"{o}fling},\ and\ \citenamefont {Bayer}}]{Kazimierczuk2015Jul}%
  \BibitemOpen
  \bibfield  {author} {\bibinfo {author} {\bibfnamefont {T.}~\bibnamefont
  {Kazimierczuk}}, \bibinfo {author} {\bibfnamefont {J.}~\bibnamefont
  {Schmutzler}}, \bibinfo {author} {\bibfnamefont {M.}~\bibnamefont
  {A{\ss}mann}}, \bibinfo {author} {\bibfnamefont {C.}~\bibnamefont
  {Schneider}}, \bibinfo {author} {\bibfnamefont {M.}~\bibnamefont {Kamp}},
  \bibinfo {author} {\bibfnamefont {S.}~\bibnamefont {H\"{o}fling}},\ and\
  \bibinfo {author} {\bibfnamefont {M.}~\bibnamefont {Bayer}},\ }\bibfield
  {title} {\bibinfo {title} {{Photon-Statistics Excitation Spectroscopy of a
  Quantum-Dot Micropillar Laser}},\ }\href
  {https://doi.org/10.1103/PhysRevLett.115.027401} {\bibfield  {journal}
  {\bibinfo  {journal} {Physical Review Letters}\ }\textbf {\bibinfo {volume}
  {115}},\ \bibinfo {pages} {027401} (\bibinfo {year} {2015})}\BibitemShut
  {NoStop}%
\bibitem [{\citenamefont {Spasibko}\ \emph {et~al.}(2017)\citenamefont
  {Spasibko}, \citenamefont {Kopylov}, \citenamefont {Krutyanskiy},
  \citenamefont {Murzina}, \citenamefont {Leuchs},\ and\ \citenamefont
  {Chekhova}}]{Spasibko2017Nov}%
  \BibitemOpen
  \bibfield  {author} {\bibinfo {author} {\bibfnamefont {K.~{\relax Yu}.}\
  \bibnamefont {Spasibko}}, \bibinfo {author} {\bibfnamefont {D.~A.}\
  \bibnamefont {Kopylov}}, \bibinfo {author} {\bibfnamefont {V.~L.}\
  \bibnamefont {Krutyanskiy}}, \bibinfo {author} {\bibfnamefont {T.~V.}\
  \bibnamefont {Murzina}}, \bibinfo {author} {\bibfnamefont {G.}~\bibnamefont
  {Leuchs}},\ and\ \bibinfo {author} {\bibfnamefont {M.~V.}\ \bibnamefont
  {Chekhova}},\ }\bibfield  {title} {\bibinfo {title} {{Multiphoton Effects
  Enhanced due to Ultrafast Photon-Number Fluctuations}},\ }\href
  {https://doi.org/10.1103/PhysRevLett.119.223603} {\bibfield  {journal}
  {\bibinfo  {journal} {Physical Review Letters}\ }\textbf {\bibinfo {volume}
  {119}},\ \bibinfo {pages} {223603} (\bibinfo {year} {2017})}\BibitemShut
  {NoStop}%
\bibitem [{\citenamefont {Vasylyev}\ \emph {et~al.}(2016)\citenamefont
  {Vasylyev}, \citenamefont {Semenov},\ and\ \citenamefont
  {Vogel}}]{Vasylyev2016Aug}%
  \BibitemOpen
  \bibfield  {author} {\bibinfo {author} {\bibfnamefont {D.}~\bibnamefont
  {Vasylyev}}, \bibinfo {author} {\bibfnamefont {A.~A.}\ \bibnamefont
  {Semenov}},\ and\ \bibinfo {author} {\bibfnamefont {W.}~\bibnamefont
  {Vogel}},\ }\bibfield  {title} {\bibinfo {title} {{Atmospheric Quantum
  Channels with Weak and Strong Turbulence}},\ }\href
  {https://doi.org/10.1103/PhysRevLett.117.090501} {\bibfield  {journal}
  {\bibinfo  {journal} {Physical Review Letters}\ }\textbf {\bibinfo {volume}
  {117}},\ \bibinfo {pages} {090501} (\bibinfo {year} {2016})}\BibitemShut
  {NoStop}%
\bibitem [{\citenamefont {Dravins}\ \emph {et~al.}(1997)\citenamefont
  {Dravins}, \citenamefont {Lindegren}, \citenamefont {Mezey},\ and\
  \citenamefont {Young}}]{Dravins1997}%
  \BibitemOpen
  \bibfield  {author} {\bibinfo {author} {\bibfnamefont {D.}~\bibnamefont
  {Dravins}}, \bibinfo {author} {\bibfnamefont {L.}~\bibnamefont {Lindegren}},
  \bibinfo {author} {\bibfnamefont {E.}~\bibnamefont {Mezey}},\ and\ \bibinfo
  {author} {\bibfnamefont {A.~T.}\ \bibnamefont {Young}},\ }\bibfield  {title}
  {\bibinfo {title} {Atmospheric intensity scintillation of stars, i.
  statistical distributions and temporal properties},\ }\href
  {https://doi.org/10.1086/133872} {\bibfield  {journal} {\bibinfo  {journal}
  {Publications of the Astronomical Society of the Pacific}\ }\textbf {\bibinfo
  {volume} {109}},\ \bibinfo {pages} {173} (\bibinfo {year}
  {1997})}\BibitemShut {NoStop}%
\bibitem [{\citenamefont {Toyoshima}\ \emph {et~al.}(2011)\citenamefont
  {Toyoshima}, \citenamefont {Takenaka},\ and\ \citenamefont
  {Takayama}}]{Toyoshima2011Aug}%
  \BibitemOpen
  \bibfield  {author} {\bibinfo {author} {\bibfnamefont {M.}~\bibnamefont
  {Toyoshima}}, \bibinfo {author} {\bibfnamefont {H.}~\bibnamefont
  {Takenaka}},\ and\ \bibinfo {author} {\bibfnamefont {Y.}~\bibnamefont
  {Takayama}},\ }\bibfield  {title} {\bibinfo {title} {{Atmospheric
  turbulence-induced fading channel model for space-to-ground laser
  communications links}},\ }\href {https://doi.org/10.1364/OE.19.015965}
  {\bibfield  {journal} {\bibinfo  {journal} {Optics Express}\ }\textbf
  {\bibinfo {volume} {19}},\ \bibinfo {pages} {15965} (\bibinfo {year}
  {2011})}\BibitemShut {NoStop}%
\bibitem [{\citenamefont {Shen}\ \emph {et~al.}(2014)\citenamefont {Shen},
  \citenamefont {Yu},\ and\ \citenamefont {Fan}}]{Shen2014}%
  \BibitemOpen
  \bibfield  {author} {\bibinfo {author} {\bibfnamefont {H.}~\bibnamefont
  {Shen}}, \bibinfo {author} {\bibfnamefont {L.}~\bibnamefont {Yu}},\ and\
  \bibinfo {author} {\bibfnamefont {C.}~\bibnamefont {Fan}},\ }\bibfield
  {title} {\bibinfo {title} {Temporal spectrum of atmospheric scintillation and
  the effects of aperture averaging and time averaging},\ }\href
  {https://doi.org/10.1016/j.optcom.2014.05.039} {\bibfield  {journal}
  {\bibinfo  {journal} {Optics Communications}\ }\textbf {\bibinfo {volume}
  {330}},\ \bibinfo {pages} {160} (\bibinfo {year} {2014})}\BibitemShut
  {NoStop}%
\bibitem [{\citenamefont {Liao}\ \emph {et~al.}(2018)\citenamefont {Liao},
  \citenamefont {Cai}, \citenamefont {Handsteiner}, \citenamefont {Liu},
  \citenamefont {Yin}, \citenamefont {Zhang}, \citenamefont {Rauch},
  \citenamefont {Fink}, \citenamefont {Ren}, \citenamefont {Liu}, \citenamefont
  {Li}, \citenamefont {Shen}, \citenamefont {Cao}, \citenamefont {Li},
  \citenamefont {Wang}, \citenamefont {Huang}, \citenamefont {Deng},
  \citenamefont {Xi}, \citenamefont {Ma}, \citenamefont {Hu}, \citenamefont
  {Li}, \citenamefont {Liu}, \citenamefont {Koidl}, \citenamefont {Wang},
  \citenamefont {Chen}, \citenamefont {Wang}, \citenamefont {Steindorfer},
  \citenamefont {Kirchner}, \citenamefont {Lu}, \citenamefont {Shu},
  \citenamefont {Ursin}, \citenamefont {Scheidl}, \citenamefont {Peng},
  \citenamefont {Wang}, \citenamefont {Zeilinger},\ and\ \citenamefont
  {Pan}}]{Liao2018Jan}%
  \BibitemOpen
  \bibfield  {author} {\bibinfo {author} {\bibfnamefont {S.-K.}\ \bibnamefont
  {Liao}}, \bibinfo {author} {\bibfnamefont {W.-Q.}\ \bibnamefont {Cai}},
  \bibinfo {author} {\bibfnamefont {J.}~\bibnamefont {Handsteiner}}, \bibinfo
  {author} {\bibfnamefont {B.}~\bibnamefont {Liu}}, \bibinfo {author}
  {\bibfnamefont {J.}~\bibnamefont {Yin}}, \bibinfo {author} {\bibfnamefont
  {L.}~\bibnamefont {Zhang}}, \bibinfo {author} {\bibfnamefont
  {D.}~\bibnamefont {Rauch}}, \bibinfo {author} {\bibfnamefont
  {M.}~\bibnamefont {Fink}}, \bibinfo {author} {\bibfnamefont {J.-G.}\
  \bibnamefont {Ren}}, \bibinfo {author} {\bibfnamefont {W.-Y.}\ \bibnamefont
  {Liu}}, \bibinfo {author} {\bibfnamefont {Y.}~\bibnamefont {Li}}, \bibinfo
  {author} {\bibfnamefont {Q.}~\bibnamefont {Shen}}, \bibinfo {author}
  {\bibfnamefont {Y.}~\bibnamefont {Cao}}, \bibinfo {author} {\bibfnamefont
  {F.-Z.}\ \bibnamefont {Li}}, \bibinfo {author} {\bibfnamefont {J.-F.}\
  \bibnamefont {Wang}}, \bibinfo {author} {\bibfnamefont {Y.-M.}\ \bibnamefont
  {Huang}}, \bibinfo {author} {\bibfnamefont {L.}~\bibnamefont {Deng}},
  \bibinfo {author} {\bibfnamefont {T.}~\bibnamefont {Xi}}, \bibinfo {author}
  {\bibfnamefont {L.}~\bibnamefont {Ma}}, \bibinfo {author} {\bibfnamefont
  {T.}~\bibnamefont {Hu}}, \bibinfo {author} {\bibfnamefont {L.}~\bibnamefont
  {Li}}, \bibinfo {author} {\bibfnamefont {N.-L.}\ \bibnamefont {Liu}},
  \bibinfo {author} {\bibfnamefont {F.}~\bibnamefont {Koidl}}, \bibinfo
  {author} {\bibfnamefont {P.}~\bibnamefont {Wang}}, \bibinfo {author}
  {\bibfnamefont {Y.-A.}\ \bibnamefont {Chen}}, \bibinfo {author}
  {\bibfnamefont {X.-B.}\ \bibnamefont {Wang}}, \bibinfo {author}
  {\bibfnamefont {M.}~\bibnamefont {Steindorfer}}, \bibinfo {author}
  {\bibfnamefont {G.}~\bibnamefont {Kirchner}}, \bibinfo {author}
  {\bibfnamefont {C.-Y.}\ \bibnamefont {Lu}}, \bibinfo {author} {\bibfnamefont
  {R.}~\bibnamefont {Shu}}, \bibinfo {author} {\bibfnamefont {R.}~\bibnamefont
  {Ursin}}, \bibinfo {author} {\bibfnamefont {T.}~\bibnamefont {Scheidl}},
  \bibinfo {author} {\bibfnamefont {C.-Z.}\ \bibnamefont {Peng}}, \bibinfo
  {author} {\bibfnamefont {J.-Y.}\ \bibnamefont {Wang}}, \bibinfo {author}
  {\bibfnamefont {A.}~\bibnamefont {Zeilinger}},\ and\ \bibinfo {author}
  {\bibfnamefont {J.-W.}\ \bibnamefont {Pan}},\ }\bibfield  {title} {\bibinfo
  {title} {{Satellite-Relayed Intercontinental Quantum Network}},\ }\href
  {https://doi.org/10.1103/PhysRevLett.120.030501} {\bibfield  {journal}
  {\bibinfo  {journal} {Physical Review Letters}\ }\textbf {\bibinfo {volume}
  {120}},\ \bibinfo {pages} {030501} (\bibinfo {year} {2018})}\BibitemShut
  {NoStop}%
\bibitem [{\citenamefont {Yin}\ \emph {et~al.}(2020)\citenamefont {Yin},
  \citenamefont {Li}, \citenamefont {Liao}, \citenamefont {Yang}, \citenamefont
  {Cao}, \citenamefont {Zhang}, \citenamefont {Ren}, \citenamefont {Cai},
  \citenamefont {Liu}, \citenamefont {Li}, \citenamefont {Shu}, \citenamefont
  {Huang}, \citenamefont {Deng}, \citenamefont {Li}, \citenamefont {Zhang},
  \citenamefont {Liu}, \citenamefont {Chen}, \citenamefont {Lu}, \citenamefont
  {Wang}, \citenamefont {Xu}, \citenamefont {Wang}, \citenamefont {Peng},
  \citenamefont {Ekert},\ and\ \citenamefont {Pan}}]{Yin2020Jun}%
  \BibitemOpen
  \bibfield  {author} {\bibinfo {author} {\bibfnamefont {J.}~\bibnamefont
  {Yin}}, \bibinfo {author} {\bibfnamefont {Y.-H.}\ \bibnamefont {Li}},
  \bibinfo {author} {\bibfnamefont {S.-K.}\ \bibnamefont {Liao}}, \bibinfo
  {author} {\bibfnamefont {M.}~\bibnamefont {Yang}}, \bibinfo {author}
  {\bibfnamefont {Y.}~\bibnamefont {Cao}}, \bibinfo {author} {\bibfnamefont
  {L.}~\bibnamefont {Zhang}}, \bibinfo {author} {\bibfnamefont {J.-G.}\
  \bibnamefont {Ren}}, \bibinfo {author} {\bibfnamefont {W.-Q.}\ \bibnamefont
  {Cai}}, \bibinfo {author} {\bibfnamefont {W.-Y.}\ \bibnamefont {Liu}},
  \bibinfo {author} {\bibfnamefont {S.-L.}\ \bibnamefont {Li}}, \bibinfo
  {author} {\bibfnamefont {R.}~\bibnamefont {Shu}}, \bibinfo {author}
  {\bibfnamefont {Y.-M.}\ \bibnamefont {Huang}}, \bibinfo {author}
  {\bibfnamefont {L.}~\bibnamefont {Deng}}, \bibinfo {author} {\bibfnamefont
  {L.}~\bibnamefont {Li}}, \bibinfo {author} {\bibfnamefont {Q.}~\bibnamefont
  {Zhang}}, \bibinfo {author} {\bibfnamefont {N.-L.}\ \bibnamefont {Liu}},
  \bibinfo {author} {\bibfnamefont {Y.-A.}\ \bibnamefont {Chen}}, \bibinfo
  {author} {\bibfnamefont {C.-Y.}\ \bibnamefont {Lu}}, \bibinfo {author}
  {\bibfnamefont {X.-B.}\ \bibnamefont {Wang}}, \bibinfo {author}
  {\bibfnamefont {F.}~\bibnamefont {Xu}}, \bibinfo {author} {\bibfnamefont
  {J.-Y.}\ \bibnamefont {Wang}}, \bibinfo {author} {\bibfnamefont {C.-Z.}\
  \bibnamefont {Peng}}, \bibinfo {author} {\bibfnamefont {A.~K.}\ \bibnamefont
  {Ekert}},\ and\ \bibinfo {author} {\bibfnamefont {J.-W.}\ \bibnamefont
  {Pan}},\ }\bibfield  {title} {\bibinfo {title} {{Entanglement-based secure
  quantum cryptography over 1,120 kilometres}},\ }\href
  {https://doi.org/10.1038/s41586-020-2401-y} {\bibfield  {journal} {\bibinfo
  {journal} {Nature}\ }\textbf {\bibinfo {volume} {582}},\ \bibinfo {pages}
  {501} (\bibinfo {year} {2020})}\BibitemShut {NoStop}%
\bibitem [{\citenamefont {Usenko}\ \emph {et~al.}(2012)\citenamefont {Usenko},
  \citenamefont {Heim}, \citenamefont {Peuntinger}, \citenamefont {Wittmann},
  \citenamefont {Marquardt}, \citenamefont {Leuchs},\ and\ \citenamefont
  {Filip}}]{Usenko2012Sep}%
  \BibitemOpen
  \bibfield  {author} {\bibinfo {author} {\bibfnamefont {V.~C.}\ \bibnamefont
  {Usenko}}, \bibinfo {author} {\bibfnamefont {B.}~\bibnamefont {Heim}},
  \bibinfo {author} {\bibfnamefont {C.}~\bibnamefont {Peuntinger}}, \bibinfo
  {author} {\bibfnamefont {C.}~\bibnamefont {Wittmann}}, \bibinfo {author}
  {\bibfnamefont {C.}~\bibnamefont {Marquardt}}, \bibinfo {author}
  {\bibfnamefont {G.}~\bibnamefont {Leuchs}},\ and\ \bibinfo {author}
  {\bibfnamefont {R.}~\bibnamefont {Filip}},\ }\bibfield  {title} {\bibinfo
  {title} {{Entanglement of Gaussian states and the applicability to quantum
  key distribution over fading channels}},\ }\href
  {https://doi.org/10.1088/1367-2630/14/9/093048} {\bibfield  {journal}
  {\bibinfo  {journal} {New Journal of Physics}\ }\textbf {\bibinfo {volume}
  {14}},\ \bibinfo {pages} {093048} (\bibinfo {year} {2012})}\BibitemShut
  {NoStop}%
\bibitem [{\citenamefont {Gong}\ \emph {et~al.}(2018)\citenamefont {Gong},
  \citenamefont {Yang}, \citenamefont {Yong}, \citenamefont {Guan},
  \citenamefont {Shentu}, \citenamefont {Liu}, \citenamefont {Li},
  \citenamefont {Cao}, \citenamefont {Yin}, \citenamefont {Liao}, \citenamefont
  {Ren}, \citenamefont {Zhang}, \citenamefont {Peng},\ and\ \citenamefont
  {Pan}}]{Gong2018Jul}%
  \BibitemOpen
  \bibfield  {author} {\bibinfo {author} {\bibfnamefont {Y.-H.}\ \bibnamefont
  {Gong}}, \bibinfo {author} {\bibfnamefont {K.-X.}\ \bibnamefont {Yang}},
  \bibinfo {author} {\bibfnamefont {H.-L.}\ \bibnamefont {Yong}}, \bibinfo
  {author} {\bibfnamefont {J.-Y.}\ \bibnamefont {Guan}}, \bibinfo {author}
  {\bibfnamefont {G.-L.}\ \bibnamefont {Shentu}}, \bibinfo {author}
  {\bibfnamefont {C.}~\bibnamefont {Liu}}, \bibinfo {author} {\bibfnamefont
  {F.-Z.}\ \bibnamefont {Li}}, \bibinfo {author} {\bibfnamefont
  {Y.}~\bibnamefont {Cao}}, \bibinfo {author} {\bibfnamefont {J.}~\bibnamefont
  {Yin}}, \bibinfo {author} {\bibfnamefont {S.-K.}\ \bibnamefont {Liao}},
  \bibinfo {author} {\bibfnamefont {J.-G.}\ \bibnamefont {Ren}}, \bibinfo
  {author} {\bibfnamefont {Q.}~\bibnamefont {Zhang}}, \bibinfo {author}
  {\bibfnamefont {C.-Z.}\ \bibnamefont {Peng}},\ and\ \bibinfo {author}
  {\bibfnamefont {J.-W.}\ \bibnamefont {Pan}},\ }\bibfield  {title} {\bibinfo
  {title} {{Free-space quantum key distribution in urban daylight with the SPGD
  algorithm control of a deformable mirror}},\ }\href
  {https://doi.org/10.1364/OE.26.018897} {\bibfield  {journal} {\bibinfo
  {journal} {Optics Express}\ }\textbf {\bibinfo {volume} {26}},\ \bibinfo
  {pages} {18897} (\bibinfo {year} {2018})}\BibitemShut {NoStop}%
\bibitem [{\citenamefont {Ruppert}\ \emph {et~al.}(2019)\citenamefont
  {Ruppert}, \citenamefont {Peuntinger}, \citenamefont {Heim}, \citenamefont
  {G{\"{u}}nthner}, \citenamefont {Usenko}, \citenamefont {Elser},
  \citenamefont {Leuchs}, \citenamefont {Filip},\ and\ \citenamefont
  {Marquardt}}]{Ruppert2019Dec}%
  \BibitemOpen
  \bibfield  {author} {\bibinfo {author} {\bibfnamefont {L.}~\bibnamefont
  {Ruppert}}, \bibinfo {author} {\bibfnamefont {C.}~\bibnamefont {Peuntinger}},
  \bibinfo {author} {\bibfnamefont {B.}~\bibnamefont {Heim}}, \bibinfo {author}
  {\bibfnamefont {K.}~\bibnamefont {G{\"{u}}nthner}}, \bibinfo {author}
  {\bibfnamefont {V.~C.}\ \bibnamefont {Usenko}}, \bibinfo {author}
  {\bibfnamefont {D.}~\bibnamefont {Elser}}, \bibinfo {author} {\bibfnamefont
  {G.}~\bibnamefont {Leuchs}}, \bibinfo {author} {\bibfnamefont
  {R.}~\bibnamefont {Filip}},\ and\ \bibinfo {author} {\bibfnamefont
  {C.}~\bibnamefont {Marquardt}},\ }\bibfield  {title} {\bibinfo {title}
  {{Fading channel estimation for free-space continuous-variable secure quantum
  communication}},\ }\href {https://doi.org/10.1088/1367-2630/ab5dd3}
  {\bibfield  {journal} {\bibinfo  {journal} {New Journal of Physics}\ }\textbf
  {\bibinfo {volume} {21}},\ \bibinfo {pages} {123036} (\bibinfo {year}
  {2019})}\BibitemShut {NoStop}%
\bibitem [{\citenamefont {Cao}\ \emph {et~al.}(2020)\citenamefont {Cao},
  \citenamefont {Li}, \citenamefont {Yang}, \citenamefont {Jiang},
  \citenamefont {Li}, \citenamefont {Hu}, \citenamefont {Abulizi},
  \citenamefont {Li}, \citenamefont {Zhang}, \citenamefont {Sun}, \citenamefont
  {Liu}, \citenamefont {Jiang}, \citenamefont {Liao}, \citenamefont {Ren},
  \citenamefont {Li}, \citenamefont {You}, \citenamefont {Wang}, \citenamefont
  {Yin}, \citenamefont {Lu}, \citenamefont {Wang}, \citenamefont {Zhang},
  \citenamefont {Peng},\ and\ \citenamefont {Pan}}]{Cao2020Dec}%
  \BibitemOpen
  \bibfield  {author} {\bibinfo {author} {\bibfnamefont {Y.}~\bibnamefont
  {Cao}}, \bibinfo {author} {\bibfnamefont {Y.-H.}\ \bibnamefont {Li}},
  \bibinfo {author} {\bibfnamefont {K.-X.}\ \bibnamefont {Yang}}, \bibinfo
  {author} {\bibfnamefont {Y.-F.}\ \bibnamefont {Jiang}}, \bibinfo {author}
  {\bibfnamefont {S.-L.}\ \bibnamefont {Li}}, \bibinfo {author} {\bibfnamefont
  {X.-L.}\ \bibnamefont {Hu}}, \bibinfo {author} {\bibfnamefont
  {M.}~\bibnamefont {Abulizi}}, \bibinfo {author} {\bibfnamefont {C.-L.}\
  \bibnamefont {Li}}, \bibinfo {author} {\bibfnamefont {W.}~\bibnamefont
  {Zhang}}, \bibinfo {author} {\bibfnamefont {Q.-C.}\ \bibnamefont {Sun}},
  \bibinfo {author} {\bibfnamefont {W.-Y.}\ \bibnamefont {Liu}}, \bibinfo
  {author} {\bibfnamefont {X.}~\bibnamefont {Jiang}}, \bibinfo {author}
  {\bibfnamefont {S.-K.}\ \bibnamefont {Liao}}, \bibinfo {author}
  {\bibfnamefont {J.-G.}\ \bibnamefont {Ren}}, \bibinfo {author} {\bibfnamefont
  {H.}~\bibnamefont {Li}}, \bibinfo {author} {\bibfnamefont {L.}~\bibnamefont
  {You}}, \bibinfo {author} {\bibfnamefont {Z.}~\bibnamefont {Wang}}, \bibinfo
  {author} {\bibfnamefont {J.}~\bibnamefont {Yin}}, \bibinfo {author}
  {\bibfnamefont {C.-Y.}\ \bibnamefont {Lu}}, \bibinfo {author} {\bibfnamefont
  {X.-B.}\ \bibnamefont {Wang}}, \bibinfo {author} {\bibfnamefont
  {Q.}~\bibnamefont {Zhang}}, \bibinfo {author} {\bibfnamefont {C.-Z.}\
  \bibnamefont {Peng}},\ and\ \bibinfo {author} {\bibfnamefont {J.-W.}\
  \bibnamefont {Pan}},\ }\bibfield  {title} {\bibinfo {title} {{Long-Distance
  Free-Space Measurement-Device-Independent Quantum Key Distribution}},\ }\href
  {https://doi.org/10.1103/PhysRevLett.125.260503} {\bibfield  {journal}
  {\bibinfo  {journal} {Physical Review Letters}\ }\textbf {\bibinfo {volume}
  {125}},\ \bibinfo {pages} {260503} (\bibinfo {year} {2020})}\BibitemShut
  {NoStop}%
\bibitem [{\citenamefont {Lo}(2018)}]{Lo2018Sep}%
  \BibitemOpen
  \bibfield  {author} {\bibinfo {author} {\bibfnamefont {A.}~\bibnamefont
  {Lo}},\ }\bibfield  {title} {\bibinfo {title} {{Demystifying the Integrated
  Tail Probability Expectation Formula}},\ }\href
  {https://doi.org/10.1080/00031305.2018.1497541} {\bibfield  {journal}
  {\bibinfo  {journal} {American Statistician}\ }\textbf {\bibinfo {volume}
  {73}},\ \bibinfo {pages} {367} (\bibinfo {year} {2018})}\BibitemShut
  {NoStop}%
\end{thebibliography}%

\end{document}